\documentclass[symmetry,article,accept,pdftex,moreauthors]{Definitions/mdpi} 

\usepackage{subcaption}

\newcommand{\ie}{i.e.,~}
\newcommand{\eg}{e.g.,~}
\firstpage{1} 
\pubvolume{16}
\issuenum{2}
\articlenumber{194}
\pubyear{2024}
\copyrightyear{2024}
\externaleditor{\textls[-25]{Academic Editors: Igor Vladimirovich Fomin and Sergey Chervon}}
\datereceived{21 December 2023 } 
\daterevised{18 January 2024 } 
\dateaccepted{19 January 2024 } 
\datepublished{6 February 2024} 
\hreflink{https://doi.org/10.3390/\linebreak sym16020194} 

\Title{Sustaining Quasi De-Sitter Inflation with Bulk Viscosity}

\TitleCitation{Sustaining Quasi De-Sitter Inflation with Bulk Viscosity}

\Author{{Sayantani Lahiri} 
 $^{1,2}$ and Luciano Rezzolla $^{2,3,4,}$*\orcidA{}}

\AuthorCitation{Lahiri, S.; Rezzolla, L.}
\address{%
$^{1}$ \quad {Center} 
 of Applied Space Technology
  and Microgravity (ZARM), {University} of Bremen, \mbox{28359 Bremen, Germany;} {sayantani.lahiri@zarm.uni-bremen.de} 
\\
$^{2}$ \quad Institut f\"ur Theoretische Physik, Goethe Universit\"at Frankfurt,
Max-von-Laue-Str.1, \mbox{60438 Frankfurt am Main, Germany}\\
$^{3}$ \quad School of Mathematics, Trinity College Dublin, D02 PN40 Dublin, 
 Ireland\\
$^{4}$ \quad Frankfurt Institute for Advanced Studies, 
Ruth-Moufang-Str. 1, 60438 Frankfurt am Main, Germany}

\corres{Correspondence: {rezzolla@itp.uni-frankfurt.de}}

\abstract{\textcolor{black}{The de-Sitter spacetime is a maximally
    symmetric Lorentzian manifold with constant positive scalar curvature
    that plays a fundamental role in modern cosmology.}
     Here, we investigate bulk-viscosity-assisted quasi de-Sitter
    inflation, that is the period of accelerated expansion in the early
    universe during which $-\dot{H}\ll H^2$, with $H(t)$ being the Hubble
    expansion rate. We do so in the framework of a causal theory of
    relativistic hydrodynamics, which takes into account non-equilibrium
    effects associated with bulk viscosity, which may have been present as the early
    universe underwent an accelerated expansion. In this framework, the
    existence of a quasi de-Sitter universe emerges as a natural
    consequence of the presence of bulk viscosity, without requiring
    introducing additional scalar fields. As a result, the equation of
    state, determined by numerically solving the generalized
    momentum-conservation equation involving bulk viscosity pressure
    turns out to be time dependent. The transition timescale
    characterising its departure from an exact de-Sitter phase is
    intricately related to the magnitude of the bulk viscosity. We
    examine the properties of the new equation of state, as well as the
    transition timescale in the presence of bulk viscosity pressure. In
    addition, we construct a fluid description of inflation and
    demonstrate that, in the context of the causal formalism, it is
    equivalent to the scalar field theory of inflation. Our analysis also
    shows that the slow-roll conditions are realised in the
    bulk-viscosity-supported model of inflation. Finally, we examine the
    viability of our model by computing the inflationary observables,
    namely the spectral index and the tensor-to-scalar ratio of the
    curvature perturbations, and compare them with a number of different
    observations, finding good agreement in most cases.}


\begin{document}
\section{Introduction}

The inflationary paradigm plays a pivotal role in modern cosmology, which
relies on the idea that the early universe underwent
nearly exponential expansion within a very short interval of time
~\cite{PhysRevD.23.347,Linde:1981mu}. While successfully
providing explanations to all the drawbacks of the Big-Bang cosmology,
inflation also presents natural explanations concerning the large-scale
structure formation of the observable universe and the distribution of
the galaxies~\cite{Mukhanov:1981xt, Hawking1982TheDO,
  PhysRevLett.49.1110}. The~predictions of inflation associated with the
temperature anisotropies measured in the cosmic microwave background
radiation are well-supported by data sources, \eg from the BICEP2
experiment~\cite{PhysRevLett.112.241101}, Wilkinson Microwave Anisotropy
Probe (WMAP) \cite{AyonBeatoGarcia98, 2003ApJS148175S, 2007ApJS170377S,
  2011ApJS19218K}, and Planck results~\cite{Planck:2013jfk,
  akrami2020planck}.

The scalar field description is among the most-commonly adopted approaches for
addressing inflation, where a scalar field, known as the ``inflaton''
field, and a specific potential are responsible for driving the
accelerated expansion of the early universe \cite{Linde:1983gd}. In the
last few decades, extensive studies have been carried out for addressing
different facets of inflation using single scalar field models
\cite{Baumann:2009ds, bezrukov2008standard, PhysRevLett.65.3233},
multi-scalar field models \cite{Bassett:2005xm, Wands:2007bd}, and braneworld
and string theory scenarios \cite{Kachru:2003sx, Dvali:1998pa} in the
framework of both Einstein's theory and alternative theories of gravity
\cite{Starobinsky:1980te, Kanti:2015pda, Clifton:2011jh}.

An alternative approach to the scalar field description is one in which
inflation is governed by bulk viscosity pressure, providing a sufficient
amount of negative pressure, which is essential for the accelerated
expansion. In this approach, the dynamical equations are described by a
relativistic theory of non-perfect cosmological fluids
\cite{maartens1995dissipative, PhysRevD.53.5483, brevik2005dark} (for
recent works, we refer to \cite{Belinchon:2022vtk,
  Hakk:2021aav,Brevik:2020psn,Cardenas:2020exv,
  Yang:2019qza,Cruz:2018psw,Brevik:2017juz, Brevik:2017msy} and the
references therein). A well-developed and commonly employed relativistic
theory of non-perfect fluids is due to M\"uller \cite{Muller:1967zza} and
Israel and Stewart \cite{israel1976nonstationary,israel1976thermo} and is
normally referred to as the MIS theory. This phenomenological
formulation, which is hyperbolic and stable, relies on the assumption
that dissipative fluxes (in the form of bulk viscosity, shear viscosity,
heat flux, and charge flow) in the system amounts to small departures from
equilibrium \cite{hiscock1985generic, hiscock1983stability} (see also
\cite{Rezzolla_book:2013} for an introduction). Consequently, the effects of
these fluxes may be treated as perturbations to the equilibrium
quantities.

Although the MIS theory is a valid relativistic hydrodynamical
description of non-perfect fluids, it does not provide a reasonable
description when applied to the epoch of the accelerated expansion of the
early universe. This is because the basic assumptions behind the MIS
theory (small departures away from equilibrium) are no longer respected
in this epoch. More specifically, during inflation, the Strong Energy
Condition (SEC) was violated, thus allowing bulk viscosity pressure to be
comparable with the equilibrium pressure and energy density of the
cosmological fluid. As a result, the departures from the equilibrium were
large, instilling sufficiently large non-equilibrium effects, and the
universe did not remain close to equilibrium during the inflationary
period. Thus, the MIS formalism may fail to provide an accurate
description of the cosmological evolution during inflation.

Due to the lack of a well-understood theory of non-equilibrium relativistic
hydrodynamics valid during inflation, the out-of-equilibrium effects are
accommodated in a phenomenological model first put forward by Maartens
and Mendez \cite{maartens1997nonlinear}, who proposed a generalisation of the
MIS theory by taking into consideration the SEC. While preserving
causality, the approach proposed in \cite{maartens1997nonlinear} introduces a
new characteristic timescale that is non-vanishing when non-equilibrium
effects are present. As a result, the modified evolution equation of the
bulk viscosity pressure contains an additional coefficient responsible for the
far-from-equilibrium effects. There are several important advantages of
the formalism presented by Maartens and Mendez
\cite{maartens1997nonlinear}. First, the MIS theory is correctly retrieved in
the regime when the departures from the equilibrium are small. Second,
the theory ensures the validity of the second law of thermodynamics; hence,
the entropy production rate remains positive. Third, an upper bound on the
maximum value of the bulk viscosity pressure exists and is compatible with
the second law of thermodynamics.

Here, we extend the work of Maartens and Mendez
\cite{maartens1997nonlinear} by relaxing the exact de-Sitter condition so
as to assess whether a quasi de-Sitter expansion of the early universe,
\ie an inflationary scenario characterised by $-\dot{H} \ll H^2$, can be
actually sustained by the bulk viscosity. To this scope, we introduce an
\textit{effective} equation of state (EOS), where the relation between
the pressure and the energy density is no longer constant at all times,
but becomes time dependent as a result of non-equilibrium effects and a
time-dependent bulk viscous stress. However, as a new equilibrium is
reached, the EOS stops evolving, and a new constant relation emerges
between the pressure and the energy density. To quantify this behaviour,
our model is described in terms of three parameters: the coefficient of the
bulk viscosity $\bar{\zeta}$, the characteristic timescale of the non-equilibrium effects $k$,
and the parameter relating the equilibrium pressure and energy
density $w_0$ of the cosmological fluid.

In this way, we find that the quasi de-Sitter inflation is a natural
outcome of the presence of a nonzero bulk viscosity and, hence, does not
require the need for an inflaton field. In particular, the universe
always exhibits departure from the exact de-Sitter universe, so that, as
long as the EOS is time dependent, an exact de-Sitter universe cannot
exist. The evolution of the transition timescale of the modified EOS from
the exact de-Sitter solution is found to depend on the variation of each
of the three aforementioned parameters of the model. As an example, we
find that this transition timescale during the quasi de-Sitter expansion
is inversely proportional to the magnitude of the bulk viscosity. On the
other hand, the timescale over which the non-equilibrium effects are present
is larger than previously found, even though they do not have a
significant impact in altering the qualitative behaviour of the equation of
the state.
  
Furthermore, our analysis reveals that, as the universe evolves during
inflation, the absolute magnitude of the bulk viscosity pressure
diminishes before settling to a small and nearly constant value. We
determined admissible regions of the parameters of the model under the
quasi de Sitter conditions, and upon comparing with the exact de-Sitter
case, we find that the admissible parameter region of one of the
parameters ($w_0$; see below) is considerably larger than
what has been computed so far. Finally, as a test of our model, we compute the
inflationary observables, \ie the spectral index and the tensor-to-scalar
ratio of the curvature perturbations, and compare them with the
observations. In this way, we find that, while our predictions are
compatible with non-Planck-based observations, they are in tension based
on the Planck data.

The paper is organised as follows. Section \ref{sec-2} provides an
overview on the bulk-viscosity-driven inflation and is devoted to
developing the mathematical background for the paper. In Section
\ref{sec-3}, the numerical results are presented, which were obtained by
solving the generalized momentum-conservation equation in the presence of
bulk viscous stress under quasi de-Sitter conditions. In Section
\ref{sec-4}, we discuss the implications of the numerical results and
provide an estimate of the magnitude of the bulk viscosity coefficient
compatible with the current results of the inflationary variables. Finally,
the conclusions and outlook are reported in Section \ref{sec-5}.

\section{Bulk-Viscosity Driven Quasi De-Sitter Inflation: Mathematical~Background}
\label{sec-2}

To study the inflationary scenario in the presence of the bulk viscosity, let us
consider the homogeneous, isotropic, spatially flat
Friedmann--Lema\^{i}tre--Robertson--Walker (FLRW) metric as follows:
\begin{equation}
  ds^2= -dt^2 + a^2(t)\delta_{ij}dx^{i}dx^{j}\,,   \qquad \qquad (i=x,y,z)
  \label{metric}
\end{equation}   
where $t$ is the cosmological time comoving with the expansion, which is
the proper time measured by a free-falling observer and $a(t)$ is the
scale-factor that describes the cosmological evolution of the
universe. \textcolor{black}{Hereafter, we will adopt units in which $8\,
  \pi G= c =1$, with~$G$ being the gravitational constant and $c$ the
  speed of light.}

To study inflation in the presence of the bulk viscous stress, let us consider the
Eckart frame in which a non-perfect relativistic fluid is subjected to
heat flux, bulk, and shear viscous stresses. Since the shear stress
identically vanishes in the isotropic FLRW spacetime, the general form of
the stress energy--momentum tensor of the cosmological fluid in terms of
the heat flux $q^{\mu}$ and the bulk viscosity pressure $\Pi$ is given by
\cite{Rezzolla_book:2013}
\begin{eqnarray}
T^{\mu \nu}&=&e u^{\mu} u^{\nu} + (p+\Pi) h^{\mu \nu} +
q^{\mu}u^{\nu}+q^{\nu}u^{\mu}\,, \label{EM1}
\end{eqnarray}
where $e$ and $p$ are, respectively, the local equilibrium energy density
and the fluid pressure, while $h^{\mu\nu}:=g^{\mu\nu}+u^{\mu}u^{\nu}$ is
the projection tensor orthogonal to the four-velocity $u^{\mu}$ and
obviously satisfies the condition $h_{\mu\nu}u^{\nu}=0$
\cite{Rezzolla_book:2013}. As is customary in cosmology, we adopted a frame
comoving with the fluid, so that the four-velocity of the fluid in such a
frame is simply given by $u^{\mu}=(1,0,0,0)$ and the four-acceleration
$a^{\mu}:=u^{\beta}\nabla_{\beta}u^{\mu}$ vanishes identically. The
total effective pressure $p_\mathrm{{eff}}$ of the fluid in the presence of the bulk scalar stress $\Pi$ is expressed as
\begin{equation}
p_\mathrm{{eff}}=p+\Pi\,.
\end{equation} 
{In} 
terms of the energy--momentum tensor, the energy density, the effective
pressure, and the heat flux can be, respectively, expressed in the following
way:
\begin{eqnarray}
  \label{eq:endens}
  e&=&T_{\alpha \beta}u^{\alpha} u^{\beta}\,, \\
  \label{eq:pressure}
  p_\mathrm{{eff}}&=& p+\Pi =
  \frac{1}{3} T_{\alpha \beta} h^{\alpha \beta}\,, \\
  \label{eq:heatflux}
  q^{\mu}&=&-K({\cal D}^{\mu}_{\bot}T +T a^{\mu})\,.
\end{eqnarray}
{Here} ${\cal D}^{\mu}_{\bot}=h^{\mu \alpha}\nabla_{\alpha}$ is the
covariant derivative orthogonal to the four-velocity, $K$ is the
coefficient of the thermal conductivity, and $T$ is the local
temperature of the fluid. Equation \eqref{eq:heatflux} in the comoving
frame ($a^{\mu}=0$) of a homogeneous and isotropic universe (giving
  rise to a constant temperature) shows that the heat flux is trivially
zero \ie $u^{\mu}\nabla_{\mu}T=0$ at the background level and will
not be considered further here (see, however, \cite{Maartens:1997pa} for an
example in which the inflationary solutions of the early universe driven
by heat flux are studied in the context of the MIS formulation). Using
Equations~(\ref{metric}) and (\ref{EM1}), the Friedmann equations (\ie the
temporal and the spatial components of the Einstein equations) are given
by
\begin{eqnarray}
  3H^2 &=& \kappa^2e\,, \hspace{25mm}   \label{EE1} \\[2mm]
 2 \dot{H}+3H^2 &=&-\kappa^2(p+\Pi)\,.\hspace{11mm}  \label{EE2}
\end{eqnarray}
where \textcolor{black}{$\kappa^2:= 8\pi G = 1$}. The~Hubble function is
defined as
\begin{equation}
H(t):=\frac{\dot{a}}{a}\,,
\end{equation}
where the ``dot'' will be employed hereafter to indicate a derivative
with respect to the coordinate time $t$. From~Equations~(\ref{EE1}) and
(\ref{EE2}) we obtain that
\begin{equation}
\dot{H}=-\displaystyle \frac{\kappa^2}{2}(e+p+\Pi)\,, \label{EE3}
\end{equation}
\textls[-15]{while the the four-divergence of the energy-momentum tensor leads to the
energy-conservation} equation
\begin{eqnarray}
\dot{e}+3H(e+p+\Pi)=0\,. \label{CE1}
\end{eqnarray}

Let us now consider that $e$ and $p$ are related by a barotropic EOS
given by
\begin{equation}
  \label{eq:EOS}
  p=w_0 e\,,
\end{equation}
where $w_0$ is a constant, known as the EOS parameter, and expected to
vary in the range $-1\leq w_0 \leq 1$. The range of $w_0$ can be set from
the validity/violation of a series of energy conditions, namely the Weak
Energy Condition (WEC), Null Energy Condition (NEC), Dominant Energy
Condition (DEC), and Strong Energy Condition (SEC). For an ideal
cosmological fluid characterised by its energy density $e$ and pressure
$p$, the WEC predicts $e \geq 0$ and $e+p \geq 0$, which shows $w_0 \geq
-1$. The NEC also gives rise to the condition $w_0 \geq -1$ and $e + p \geq 0 $. 
The DEC shows that $e \geq \mid p\mid$, implying $w_0\leq 1$. On the other hand,
the SEC given by $e+3p \geq 0$ implies $w_0 \geq -1/3$. The fact that the
early universe underwent a phase of accelerated expansion results in
the violation of the SEC, i.e., $e+3p \leq 0$ or $w_0 \leq -1/3$. In
particular, the exponential expansion of the universe is generated with
the EOS $p=-e$ or $w_0=-1$. Now, in the presence of the bulk viscosity, 
using Equations~(\ref{EE1}) and~(\ref{EE2}), Equation~(\ref{EE2}) leads to
\begin{eqnarray}
  \frac{\ddot{a}}{a}&=&-\frac{\kappa^2}{6}[e+3 e \,w_{0} +3\Pi]\,.
  \label{eq-I}
  \end{eqnarray}
{The} left-hand side of Equation~(\ref{eq-I}) can also be expressed as
\begin{eqnarray}
  \frac{\ddot{a}}{a}=\dot{H}+H^2=H^2(1-\epsilon_{\rm H})\,, \qquad
  \,, \label{eq-II}
\end{eqnarray}
where $\epsilon_{\rm H}=-{\dot{H}}/{H^2}$ is also known as the slow-roll
parameter (or first Hubble-flow). Since the accelerated
expansion requires $\ddot{a}\,\textgreater \,0$, during~inflation
the following conditions are always satisfied from Equations~(\ref{eq-I}) and (\ref{eq-II})
\begin{eqnarray}
 1+3w_{0}+\frac{3\Pi}{e}  \, \textless \,0\,,\\
\qquad  0 \leq \epsilon_{\rm H}\, \textless \,1\,.
\end{eqnarray}

We note that $\dot{H}=0$ corresponds to the ``exact de-Sitter''
expansion, for~which $\epsilon_{\rm H}=0$ and $a(t) \propto e^{H_0 t}$,
where $H_0$ is a constant. The~moment corresponding to $\epsilon_{\rm
  H}=1$ is defined as the exit from inflation when $\ddot{a}=0$ and the
acceleration of the universe finally comes to a~halt.

Because we are interested in the presence of the bulk viscosity pressure here,
we introduce a new EOS, which we denote as the \textit{effective} EOS, defined in the following way:
\begin{eqnarray}
w_\mathrm{{eff}} := \frac{p_\mathrm{{eff}}}{e}&=& w_{0}+\displaystyle
\frac{\Pi}{e}\,.
\label{weff-def}
\end{eqnarray}
{Note} also that $w_\mathrm{{eff}}$ is a time-dependent function, which,
using Equations~(\ref{EE3}) and~(\ref{weff-def}), can be re-expressed in
terms of the Hubble function via an ``effective EOS parameter'':
%
\begin{eqnarray}
  w_\mathrm{{eff}} = - \left(1+
  \frac{2}{3}\frac{\dot{H}}{H^2}\right)\,. \label{eq-20a}
  \end{eqnarray}
{Hence}, so long as the condition
%
\begin{equation}
  -\frac{2}{3} \left(\frac{\dot{H}}{H^2}\right) \ll1\,,
\end{equation}
is valid, $w_\mathrm{{eff}} \gtrsim -1$ remains true and the
early universe undergoes a \textit{nearly} exponential expansion
supported by the bulk viscosity. On~the other hand, in~the absence of the
bulk-viscosity pressure Equation~(\ref{weff-def}) reduces to
\begin{equation}
  w_\mathrm{{eff}}=w_{0}\,,
\end{equation}
and becomes a constant. The~universe is then filled with a perfect fluid
and the corresponding pressure and the energy density are related by the
EOS-parameter $w_0$.

\subsection{Examining the Energy~Conditions}

In what follows, we analyse the energy conditions during the accelerated
expansion of the universe in the presence of the bulk viscosity pressure. We
start by recalling that the SEC is given by
\begin{eqnarray}
  T_{\alpha \beta}u^{\alpha}u^{\beta}-\frac{1}{2}T u^{\alpha} u_{\alpha} \geq 0\,, 
\end{eqnarray}
which can be alternatively expressed as
\begin{eqnarray}
   \qquad R_{\alpha \beta} u^{\alpha} u^{\beta}  \geq 0\,. \label{eq-21} 
  \end{eqnarray}
{In} an FLRW spacetime, Equation~(\ref{eq-21}) then reduces to
\begin{equation}
  \frac{\ddot{a}}{a} \leq 0\,,
  \end{equation}
thus highlighting that the SEC must be violated to satisfy the condition
$\ddot{a} \,\textgreater \, 0$ during the inflationary phase. On~the
other hand, the~Weak Energy Condition (WEC) and the Null Energy Condition
(NEC), respectively, give rise to the following conditions:
\begin{eqnarray}
e &\geq &0 \quad \Rightarrow \quad \frac{\dot{a}^2}{a^2}\geq 0\,, \\[1mm]
e+ p+ \Pi &\geq&0 \quad \Rightarrow \quad
-\frac{\ddot{a}}{a}+\frac{\dot{a}^2}{a^2}\geq 0\,.
\end{eqnarray}
{The} WEC remains valid because the energy density of the viscous
cosmological fluid is positive. Furthermore, since we will not consider
inflationary models corresponding to $w_\mathrm{{eff}}\, \textless \,-1$,
the NEC is also satisfied, and $\dot{H}$ remains negative during the
quasi de-Sitter expansion.

Next, we recall that the relativistic theories of non-perfect fluids
proposed by Eckart and Landau lead to dynamically unstable equilibrium
states under linear perturbations and do not give rise to hyperbolic
equations of motion that result in a violation of causality
\cite{hiscock1985generic}. To counter these drawbacks, one of the
best-developed and -studied approaches towards constructing a causal
theory of the relativistic hydrodynamics of non-perfect fluids is the MIS
formalism \cite{israel1976nonstationary, israel1976thermo}. 
This approach makes use of the second-order gradients of the hydrodynamical variables
and appropriately introduces relaxation time transport coefficients
corresponding to all the dissipative quantities. Notwithstanding the
complications brought about by this approach (see \cite{Chabanov2021} and the
references therein), the MIS formulation is able to remove all the
drawbacks of the first-order theories of the relativistic hydrodynamics of
non-perfect fluids proposed by Eckart and Landau
\cite{hiscock1985generic, hiscock1983stability}. We note that possible
alternatives to the MIS formulation have recently been proposed
addressing the existence of the hyperbolicity and causality of the hydrodynamical
theory of relativistic non-perfect fluids \cite{Bemfica:2019cop,
  Kovtun:2019hdm}; while these formulations are interesting and deserve
future attention, they will not be employed here, where we instead focus
on the MIS formalism.

The most-crucial assumption of the MIS formalism is that it considers
regimes that are near-equilibrium, that is where all the
dissipative flux quantities are small compared to the equilibrium fluid
variables. In the context of the present study, this implies that the
bulk viscosity pressure $\Pi$ must be smaller when compared to the
equilibrium fluid pressure, \ie $\Pi \ll p$, where the equilibrium cosmological
fluid pressure is always non-negative, i.e., $p>0$. For an accelerated expansion
of the early universe, Equation~(\ref{eq-I}), however, indicates that
$e+3(p+\Pi)<0 $, which, in turn, implies
\begin{eqnarray}
  -\Pi& \textgreater& p+\frac{1}{3} e\,. \label{eq-23}
\end{eqnarray} 
{Thus}, the violation of the SEC leads to a scenario where $\Pi$ is
actually greater than the equilibrium fluid variables, and the validity of the MIS theory is, therefore, put to question for studying the inflationary regime in the
presence of bulk viscosity.

Given these considerations, we adopt here the phenomenological approach
prescribed by Maartens and Mendez \cite{maartens1997nonlinear}, which not
only incorporates the out-of-equilibrium effects by providing a
minimalist modification to the MIS approach via the introduction of a
characteristic scale $\tau_{\ast}$, but also ensures that the MIS
approach is recovered in the appropriate limit. In the context of the MIS
formalism, the positivity of the entropy production leads to the following
relations \cite{Maartens:1997pa}:
\begin{eqnarray}
 \label{eq:Pichi_old}
  \Pi&=& -\zeta \chi\,,  \\
  \chi&=& 3H +
  \frac{\tau_{_{\Pi}}}{\zeta} \dot{\Pi}+\frac{\tau_{_{\Pi}}}{2 \zeta}\Pi
  \left[3H+\frac{\dot{\tau_{_{\Pi}}}}{\tau_{_{\Pi}}}-\frac{\dot{\zeta}}{\zeta}-
    \frac{\dot{T}}{T}\right]\,, \label{chi}
 \end{eqnarray}
where $\zeta$ and $\tau_{_{\Pi}}$ are the bulk viscosity and the
relaxation time coefficients, respectively (note that $\zeta$ has the
{dimensions of the inverse of mass}). To include far-fro-
equilibrium effects, the bulk viscosity pressure \eqref{eq:Pichi_old} is
modified in a nonlinear manner as \cite{maartens1997nonlinear}
\begin{eqnarray}
  \label{non-linear-pi}
  \Pi= - \frac{\zeta  \chi}{1+\tau_{\ast}\chi}\,. 
\end{eqnarray}
where $\tau_{\ast}\geq 0$ is the timescale of the onset of the
non-equilibrium effects which become significant under the condition $\chi
\gtrsim \tau_{\ast} ^{-1}$. Clearly, the~MIS approach is retrieved when
$\chi$ is small, \ie if $\chi \ll \tau_{\ast} ^{-1}$. On~the other hand,
in the limit $\chi \rightarrow \infty$, the expression~\eqref{non-linear-pi}
ensures that $-\Pi$ does not become arbitrarily large but is bounded by
the condition, $-\Pi\leq {\zeta}/{\tau_{\ast}}$. Additionally, the~second
law of thermodynamics for out-of-equilibrium systems implies that the entropy production
should always be non-negative, namely
\begin{equation}
\nabla_{\alpha} S^{\alpha} \geq 0\,.
\end{equation}
{Since} the MIS limit is recovered with $\tau_{*}=0$, the~four-divergence
of the non-equilibrium entropy in that case is given by
~\cite{maartens1997nonlinear}
\begin{equation}
\nabla_{\alpha} S^{\alpha} = \frac{\Pi^2}{T \zeta} \geq 0 \,,
\end{equation}
which implies $\zeta>0$. In~other words, to~guarantee that a smooth
$\tau_{*} \rightarrow 0$ limit always exits, the~condition $\zeta >0$
needs to be~considered.

Using now Equations~(\ref{chi}) and (\ref{non-linear-pi}), the~momentum-conservation equation of the bulk-viscosity pressure becomes
\vspace{-10pt}
\begin{adjustwidth}{-\extralength}{0cm}
\centering 
\begin{equation}
 \tau_{_{\Pi}} \dot{\Pi}\left(1+\frac{\tau_{\ast}}{\zeta} \Pi\right)+
 \Pi(1+3 H \tau_{\ast})=-3\zeta H -\displaystyle \frac{1}{2} \tau_{_{\Pi}}
 \Pi \left(3H
 +\frac{\dot{\tau_{_{\Pi}}}}{\tau_{_{\Pi}}}-\frac{\dot{\zeta}}{\zeta} -
 \frac{\dot{T}}{T}\right)\left(1+\frac{\tau_{\ast}}{\zeta}\Pi\right)\,.
 \label{master-eqn}
\end{equation}
\end{adjustwidth}
which recovers the MIS equivalent when $\tau_{*}=0$. 

Of course, the~phenomenological prescription~\eqref{non-linear-pi} does
not provide a way to determine the timescale $\tau_{*}$, which should in
principle be estimated from microscopic considerations. Once again, we
follow the prescription proposed by Maartens and Mendez
~\cite{maartens1997nonlinear}, which simply relates the relaxation time to the characteristic timescale, \ie
\begin{equation}
  \label{eq:taustar}
  \tau_{\ast}= k^2 \tau_{_{\Pi}}\,,
\end{equation} 
where $k$ is a dimensionless proportionality constant. Let us now
consider the ansatz that $\zeta \propto \sqrt{e}$, from~which we can set
\begin{equation}
  \zeta = \sqrt{3}\,\bar{\zeta}\,\frac{H}{\kappa}\,, \label{eq-7}
\end{equation}
with $\bar{\zeta}$ a proportionality constant. \textcolor{black}{Since 
  $\zeta$ has the dimensions of an inverse of a mass, using Equation~(\ref{eq-7}),
  the proportionality constant $\bar{\zeta}$ will have the same
  dimensions}. In~the absence of a heat flux and shear viscosity, the~relation between the relaxation-time coefficient and $\zeta$ is given by
~\cite{maartens1997nonlinear}
\begin{equation}
  \tau_{_{\Pi}}=\displaystyle \frac{\zeta}{e(1 +w_0)v_{b}^2 }\,, \label{eq-8}
  \end{equation}
where $v_{b}$ is the time-independent bulk viscosity (non-adiabatic)
contribution to the sound speed $V_{s}$, namely, $V_{s}^2 := c_s^2 +
v_{b}^2$, with~$c_s$ the adiabatic contribution defined as
\begin{equation}
  c_{s}^2: = \displaystyle \frac{\partial p}{\partial e} =w_0\,, \label{eq-9}
\end{equation}
where the last equality applies in the specific case of the barotropic
EOS~\eqref{eq:EOS} (see Refs.~\cite{hiscock1983stability, Maartens1996}
for a detailed discussion of the sound speed in the presence of
non-adiabatic effects). Since the sound speed cannot exceed unity (the
speed of light), \ie $w_0 + v_{b}^2 \leq 1$, the~relaxation-time
coefficient becomes
\begin{equation}
  \tau_{_{\Pi}} \geq \displaystyle \frac{\zeta}{e(1-w_0^2)} =
  \displaystyle \frac{\kappa \bar{\zeta}}{\sqrt{3}H(1-w_0^2)}\,,
  \label{eq-10}
\end{equation}
where the second expression is obtained by substituting Equation~(\ref{eq-7})
and Equation~(\ref{EE1}) in the first expression. Let us now consider the
local temperature to be a simple function of the energy density, \ie
$T=T(e)$, and~the integrability conditions of the second law of
thermodynamics (Gibbs equation)
\begin{eqnarray}
  TdS&=&(p +e)d\left(\frac{1}{n}\right)+\displaystyle \frac{1}{n}de\,,
  \\[1mm] \displaystyle \frac{\partial^2 S}{\partial e \partial
    n}&=&\displaystyle \frac{\partial^2 S}{\partial n \partial e}\,,
\end{eqnarray} 
then, from~the above two conditions, one obtains,
\begin{equation}
  T(e)\propto   \exp\left({\frac{w_0}{w_0+1}}\right)\,.
  \label{eq-11}
\end{equation}
{Using} the Friedmann Equation~(\ref{EE3}), we can express the
bulk-viscosity pressure in terms of the Hubble function as
\begin{equation}
  \Pi=-\left[2\dot{H}+3H^2(1+w_0)\right]\,.
  \label{eq-12}
  \end{equation}
{Substituting} the expressions for $\tau_{_{\Pi}}$, $T(e)$, and $v_{b}^2$,
together with Equations~(\ref{eq-7}) and (\ref{eq-9}) in
Equation~(\ref{master-eqn}), the~Friedmann equation expressing the dynamics of
the function $H=H(t)$ can be expressed as
\vspace{-10pt}
\begin{adjustwidth}{-\extralength}{0cm}
\centering 
\begin{eqnarray}
1&+&\frac{2 \left[\displaystyle\frac{2 k^2}{3} \frac{\dot{H}}{H^2}+
    (w_0+1) \left(k^2+w_0-1\right)\right] \left[3 (w_0+1)
    \displaystyle\frac{\dot{H}}{H^2}+\displaystyle\frac{\ddot{H}}{H^3}\right]}
{9\left(w_0^2-1\right)^2} \nonumber \\[2mm]
&+&\frac{
  \left(\displaystyle\frac{2\dot{H}}{3H^2} +w_0+1 \right) (w_0^2-1-\sqrt{3}k^2 \bar{\zeta})}{
  \sqrt{3}\bar{\zeta}(w_0^2-1)}\nonumber \\[2mm] &+&\frac{
  \left(\displaystyle\frac{2\dot{H}}{3H^2}+w_0+1\right)
  \left[w_0+1-\displaystyle\frac{2\dot{H}}{3H^2}(2 w_0+1) \right]
  \left[\displaystyle\frac{2 k^2}{3} \displaystyle\frac{\dot{H}}{H^2}+
    (w_0+1) \left(k^2+w_0-1\right)\right]}{2(w_0-1)^2
  (w_0+1)^3}=0\,. \label{eq-15}
\end{eqnarray}
\end{adjustwidth}
{With} an appropriate choice of the initial conditions and for a given choice
of the free parameters, namely: $\bar{\zeta}, k$, and $w_0$, the evolution
of the Hubble function $H(t)$ is determined from Equation~(\ref{eq-15}), which,
when substituted back into Equation~(\ref{eq-12}), gives the evolution of the
bulk viscosity pressure $\Pi(t)$.

An advantage of the formulation proposed in \cite{maartens1997nonlinear}
is that the bulk viscosity pressure has an upper bound given by $-\Pi
\leq{\zeta}/{\tau_{\ast}}$, which results in being fixed by the validity of the
second law of thermodynamics. With this upper-bound constraint, the
relation between $k$ and $v_b$ under the quasi de-Sitter condition on
Hubble expansion is determined as follows:
\begin{align}
 k^2    & \leq v_{b}^2\,, &           &\Rightarrow &  k^2 = \tilde{\alpha}  v_{b}^2\,, & &0\leq \tilde{\alpha} \leq 1 &\\ 
 v_{b}^2 & \leq  1-c_s^2 =1-w_0\,, &  &\Rightarrow &  v_{b}^2  = \epsilon (1-w_0)\,,  & &0\leq \epsilon \leq 1 &\,,
\end{align}
or, equivalently, 
\begin{eqnarray}
k^2=\tilde{\alpha} v_{b}^2=\alpha(1-w_0)\,,   \label{eq-16}
\end{eqnarray}
where $\alpha:=\tilde{\alpha} \epsilon$ is, again, a proportionality
constant between zero and one. The non-equilibrium characteristic
timescale $\tau_{*}$ of the non-equilibrium effects \eqref{eq:taustar} is
then given~by 
\begin{equation}
    \tau_{*} = \kappa \frac{\alpha \bar{\zeta}}{H(1+w_0)}\,.
\end{equation}
{Hence}, the~non-equilibrium characteristic timescale can be completely
parametrized by $\alpha$ for fixed values of $\bar{\zeta}$ and $w_0$.

\subsection{Exact De-Sitter Expansion within the Generalised Causal~Theory}

For completeness, we briefly examine the \textit{exact} de-Sitter
inflationary scenario generated solely by a large bulk viscosity
pressure, \ie $\Pi \gg p$ in the generalised causal framework, which was
first proposed by Maartens and Mendez \cite{maartens1997nonlinear}. The
de-Sitter expansion corresponds to $H(t)=H_0 = {\rm const.}$, with
$H_{0}\, \textgreater \,0$ and $\dot{H}(t)=0$ implying a constant EOS
given by $w_\mathrm{{eff}}=-1$. From Equations~(\ref{eq-15}) and (\ref{eq-16}),
one obtains
\begin{equation}
    -27 (w_0+1)^3 \left(-\sqrt{3} (\alpha -1) \bar{\zeta} +2 w_0^2+2 \sqrt{3}
   (\alpha -1) \bar{\zeta} w_0-2\right)=0\,.
\end{equation}
so that the bulk viscosity coefficient required to produce the
de-Sitter expansion is given by
\begin{eqnarray}
\bar{\zeta}_{\rm{dS}}=\frac{2(w_0^2-1)}{\sqrt{3}(2w_0-1)(1-\alpha)}\,. \label{eq-14}
\end{eqnarray}

\begin{figure}[H]
\includegraphics[width=0.65\textwidth]{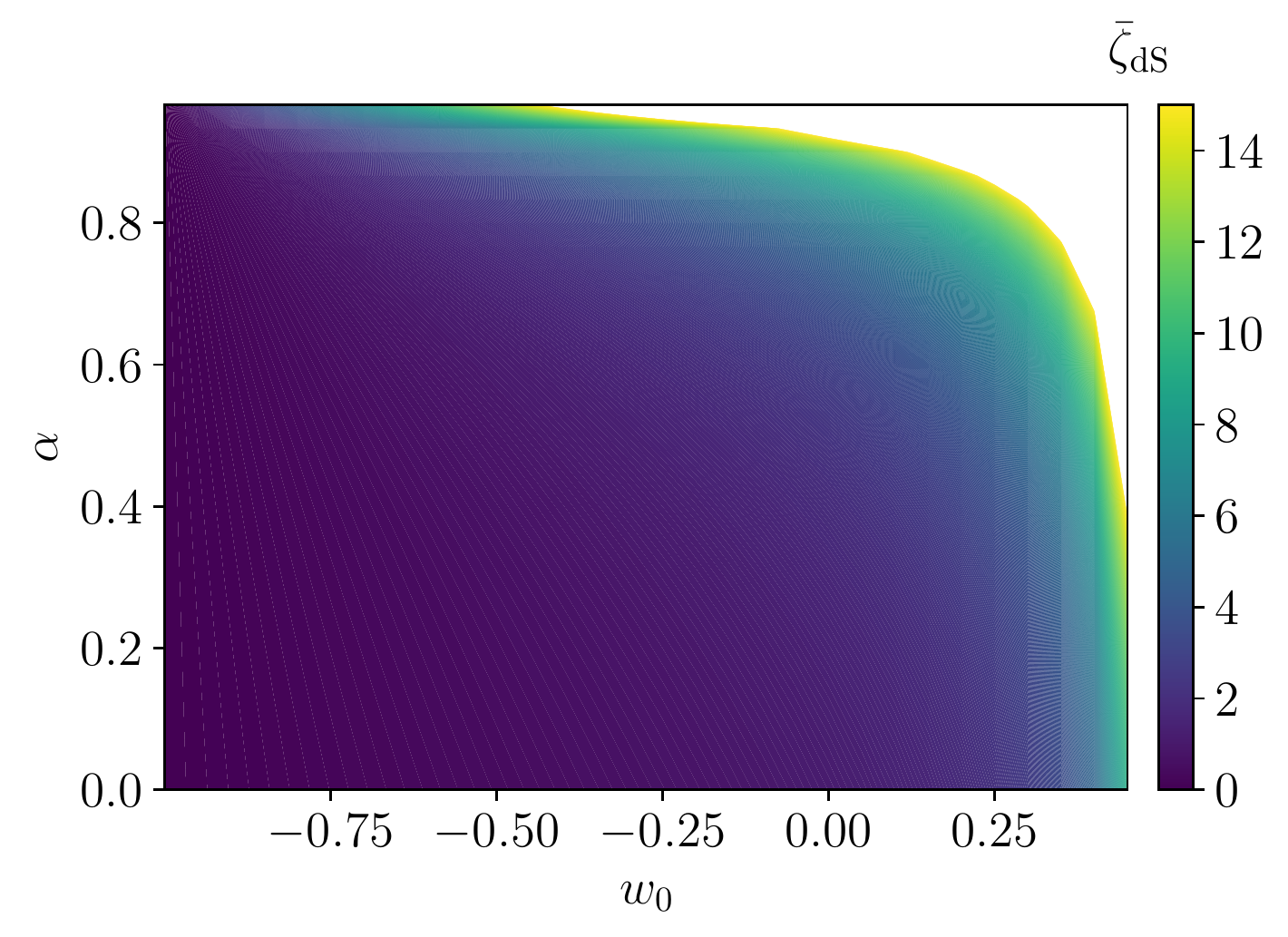}
\caption{Behaviour of the maximum value of the bulk-viscosity coefficient
  $\bar{\zeta}_{\rm{dS}}$ as a function of the two free parameters in
  the model, $\alpha$ and $w_0$ (see Equation~\eqref{zeta-max}).}
  \label{Fig-1}
\end{figure}

The variation of $\bar{\zeta}_{\rm{dS}}$ as a function $w_0$ and $\alpha$
is shown in Figure \ref{Fig-1}. Since $\bar{\zeta}_{\rm{dS}}$ has to be
positive---so as not to violate the second law of thermodynamics---and
finite, the permissible parameter space for $w_0$ becomes $-1 < w_0 <
0.5$. The solution corresponding to $w_0=-1$ gives rise to a
time-independent EOS depicting inflation driven by a perfect fluid, where the
bulk viscosity does not have any role. Thus, the \textit{exact} de-Sitter
phase arises under two distinct situations: (a) if $w_0=-1$, such that
the accelerated expansion of the early universe is as a perfect fluid; (b)
the accelerated expansion is driven by a large bulk viscosity pressure
such that the coefficient of viscosity is given by Equation~(\ref{eq-14})
\cite{maartens1997nonlinear}. In both cases, the effective EOS parameter
is time independent and is given by $w_{\rm{eff}}=-1$.

\unskip

\section{Numerical~Solutions}
\label{sec-3}

In this Section, we investigate the emergence of a quasi de-Sitter phase
supported by the bulk viscosity pressure within a causal MIS formulation. In
this scope, we numerically solve the generalized momentum conservation
Equation \eqref{master-eqn} and the effective EOS parameter
\eqref{weff-def} under the quasi de-Sitter condition so as to obtain the
evolution of the Hubble function. More specifically, substituting
Equation~(\ref{eq-16}) into Equation~(\ref{eq-15}) gives
\vspace{-10pt}
\begin{adjustwidth}{-\extralength}{0cm}
\centering 
\begin{flalign}
\left(\frac{\ddot{H}}{H^3}+3 (w_0+1)\frac{
  \dot{H}}{H^2}\right)\left(\frac{2\alpha}{3} \frac{\dot{H}}{H^2}+(\alpha
-1) (w_0+1)\right) \nonumber \\
+\,\frac{3}{2}(w_{0}^2-1)\left[\frac{\sqrt{3}}{\bar{\zeta}}
  \left(\sqrt{3} \alpha \bar{\zeta}+w_0+1\right) \left(\frac{2}{3}
  \frac{\dot{H}}{H^2}+w_0+1\right)-3(w_0+1)\right] + \nonumber \\
\frac{9}{4(w_0+1)} \left(\frac{2}{3} \frac{\dot{H}}{H^2}+
w_0+1\right) \left( w_{0}+1 -\frac{2}{3} (2 w_0+1)
\frac{\dot{H}}{H^2}\right)\left(\frac{2\alpha}{3}
\frac{\dot{H}}{H^2}+(\alpha -1) (w_0+1)\right) =0\,, \label{eq-q:1}
\end{flalign}
\end{adjustwidth}
so that the Hubble function is determined by solving Equation~(\ref{eq-q:1})
numerically for given values of $\alpha$ and $w_0$, under~the quasi
de-Sitter condition that $-\dot{H}\ll H^2$. Note that because
$w_\mathrm{eff}$ is a time-dependent function under the quasi de-Sitter
condition, we express Equation~(\ref{eq-q:1}) in terms of a first-order
ordinary differential equation in $w_\mathrm{eff}$ as follows
\begin{eqnarray} 
\frac{\dot{w}_{\rm eff}}{3H(t)} +\frac{(1+w_0+\sqrt{3}\,\alpha
  \,\bar{\zeta})(1-w_0^2)(w_\mathrm{{eff}}-w_0)}{\sqrt{3}\,
  \bar{\zeta}(1+w_0+\alpha(w_\mathrm{eff}-w_0))}
&+&\frac{(1-w_0^2)(1+w_0)}{(1+w_0+\alpha(w_\mathrm{eff}-w_0))} \nonumber\\
&-&
\frac{(w_\mathrm{{eff}}-w_0)^2}{2(1+w_0)}=0\,, \label{eq-q:2} 
\end{eqnarray}
where we have made use of the fact that
\begin{equation}
  H(t)=\frac{2}{3} \left[\int(1+w_\mathrm{eff}(t))
    dt\right]^{-1}\,. 
\end{equation}
{The} evolution of $w_{\rm{eff}}$ is determined by numerically solving
Equation~(\ref{eq-q:2}), with initial condition $w_\mathrm{eff}(t_{\rm in})
\approx -1$ (\ie $w_\mathrm{eff}-1=10^{-12}$) as expected if the universe
is nearly de-Sitter at $t=t_{\rm{ in}}$ and possesses a nonzero
bulk viscosity pressure $\Pi(t_{\rm{in}})\sim 3H_{\rm
  in}^2(w_\mathrm{eff}(t_{\rm{in}})-w_0)$, where $H_{\rm in}$ is the
Hubble constant at $t=t_{\rm in}$.

\begin{figure}[H]
  \centering
  \includegraphics[width=0.48\textwidth]{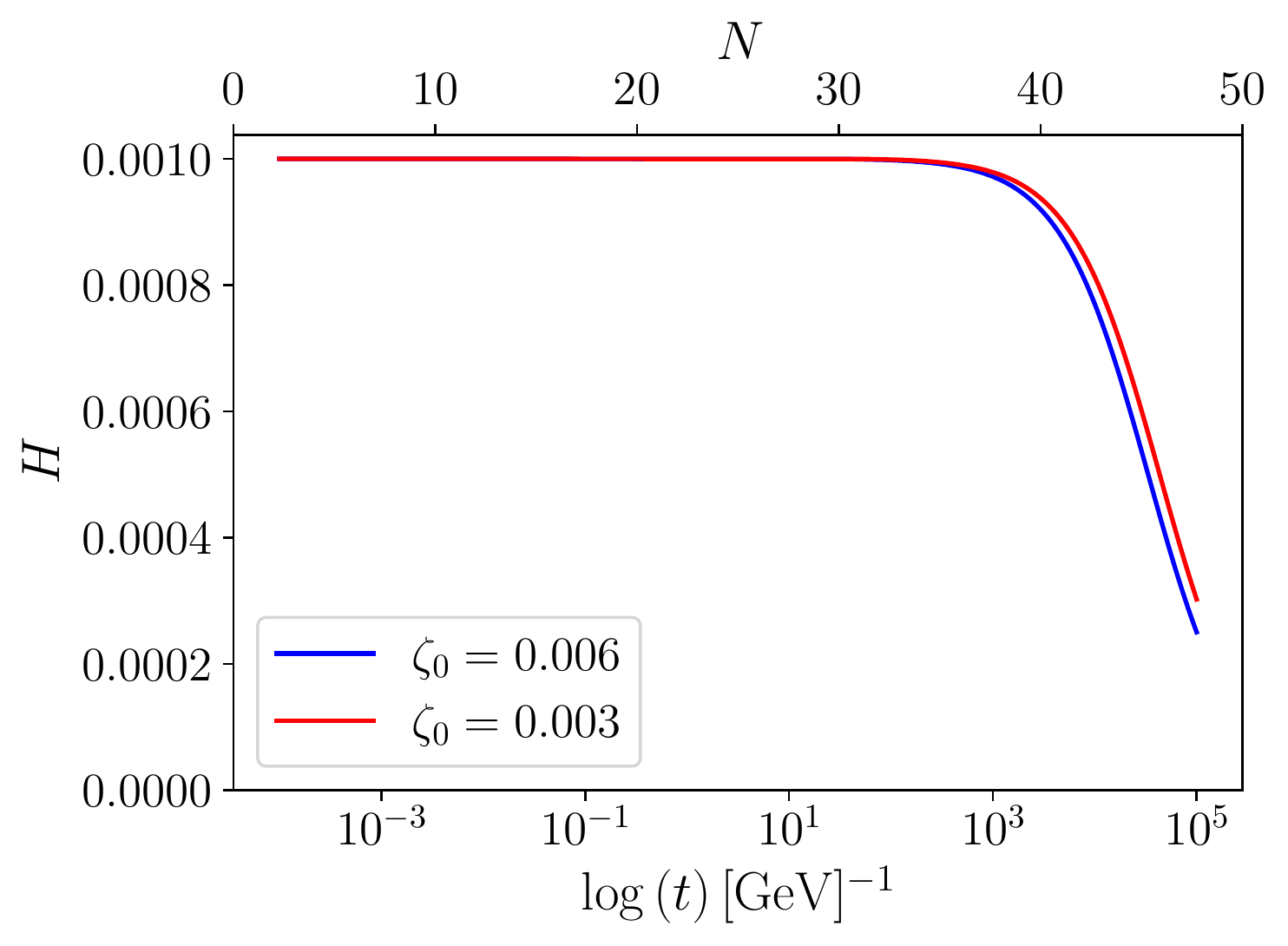}
  \hspace{3mm}
  \includegraphics[width=0.48\textwidth]{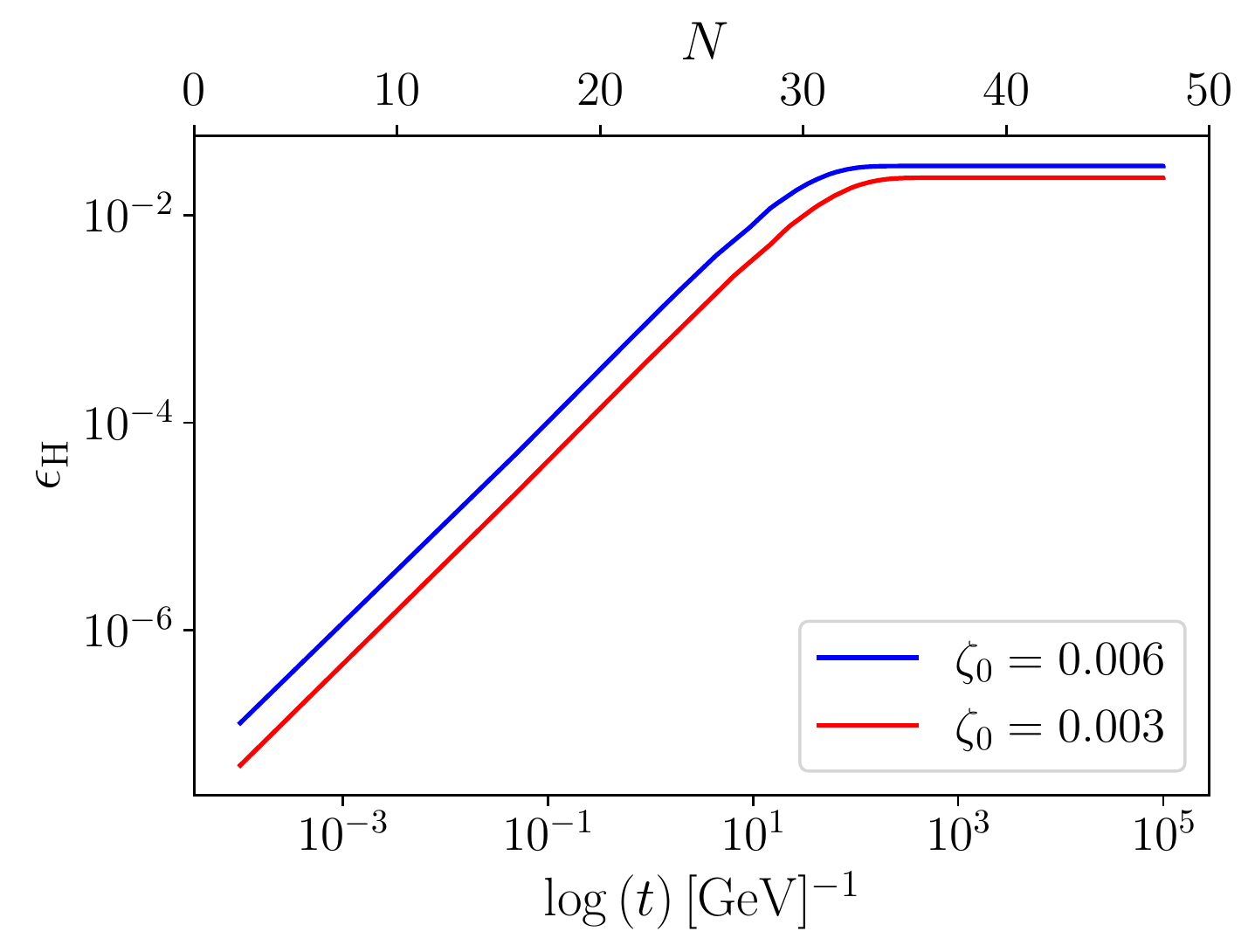}
  \caption{\textit{Left panel:} Evolution of Hubble function $H(t)$ for
    for a representative choice of $\alpha =0.1$ and $w_0 =
    -0.975$. \textit{Right panel:} The same as in the left panel but for
    the slow-roll parameter (or first Hubble-flow) parameter. In~      both panels the top horizontal axes report time in terms of the
      number of e-foldings $N$.}
\label{Fig-4}
\end{figure}

\begin{figure}[H]
  \includegraphics[width=0.5\textwidth]{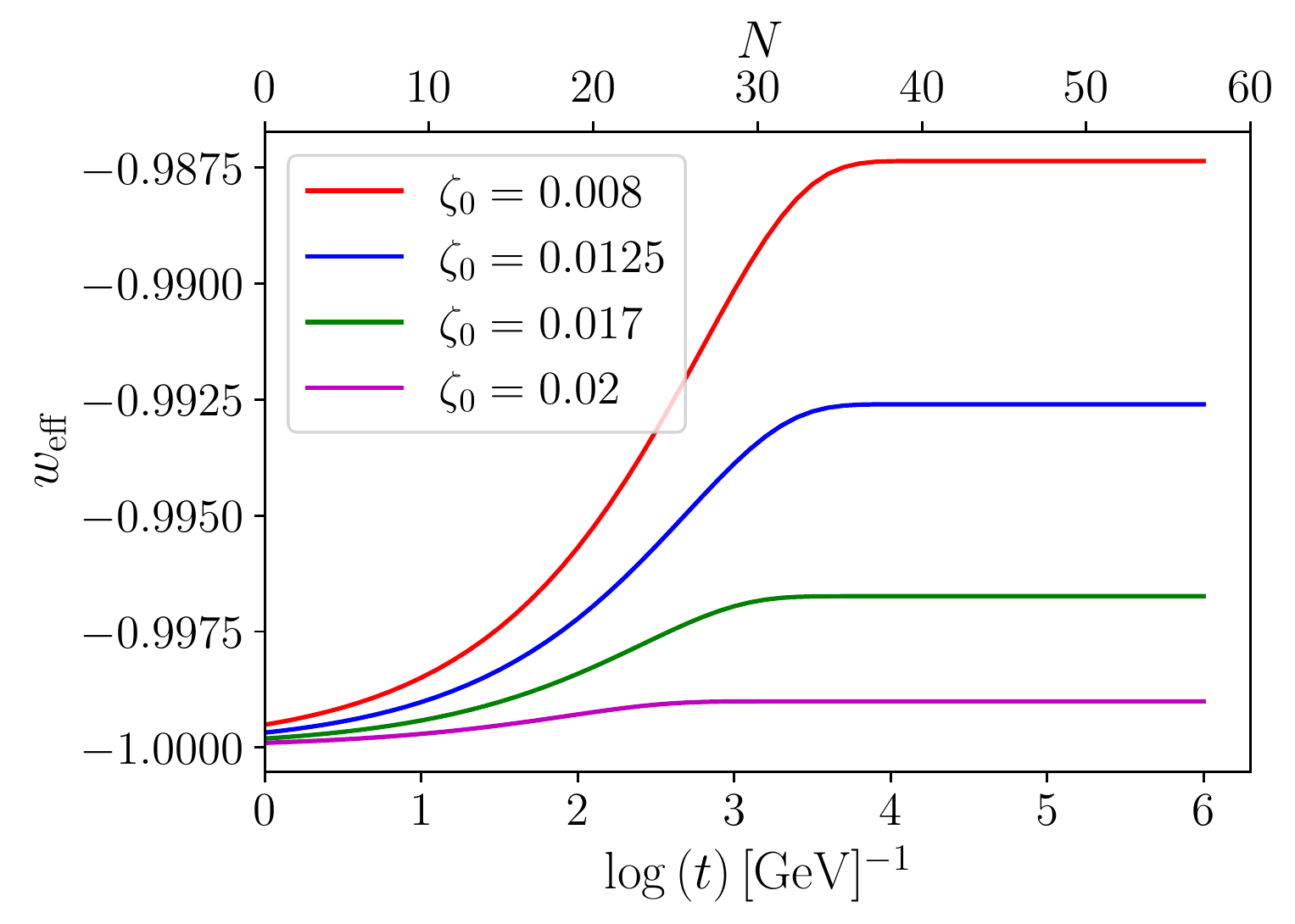}
  \includegraphics[width=0.5\textwidth]{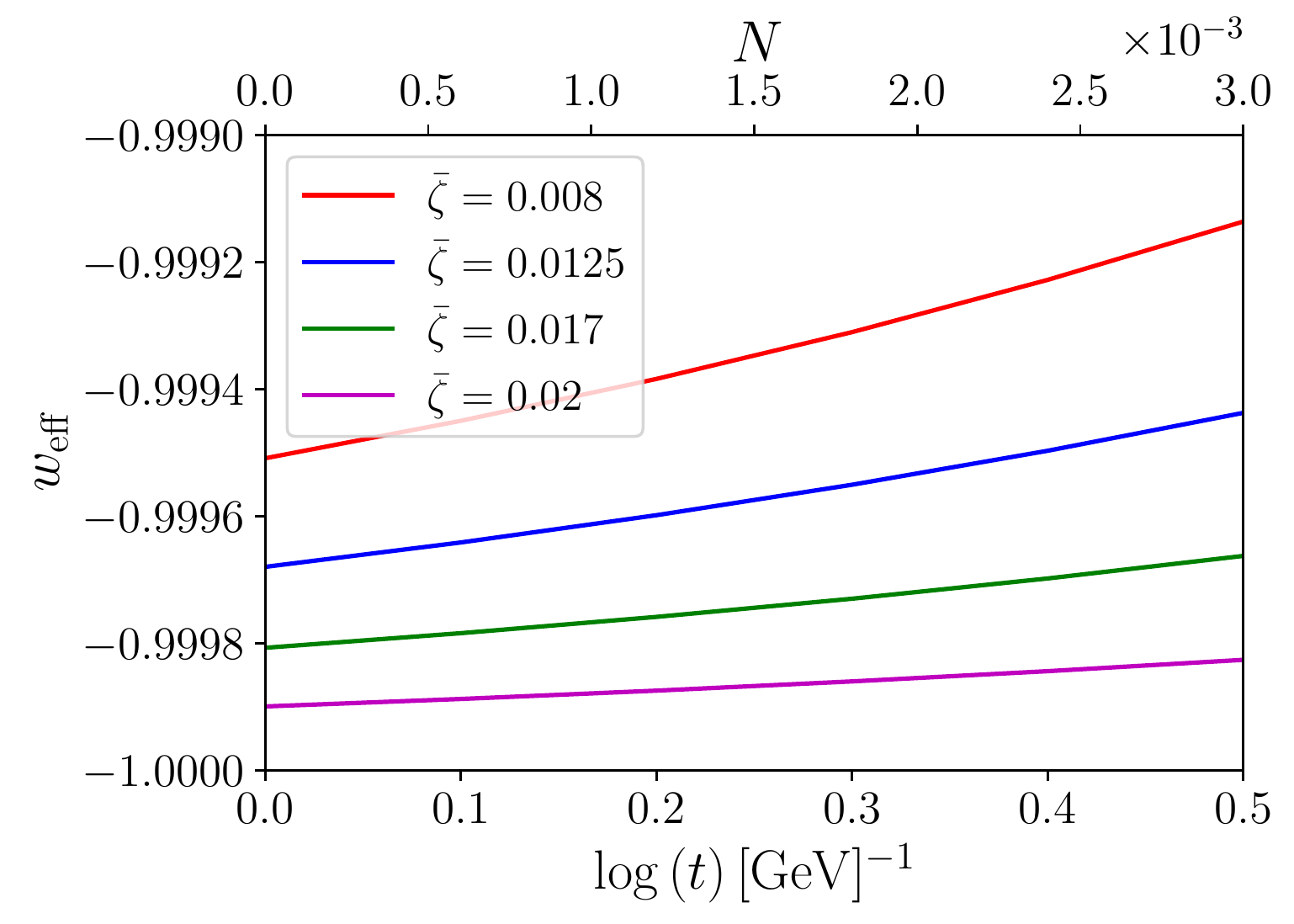} 
  \includegraphics[width=0.5\textwidth]{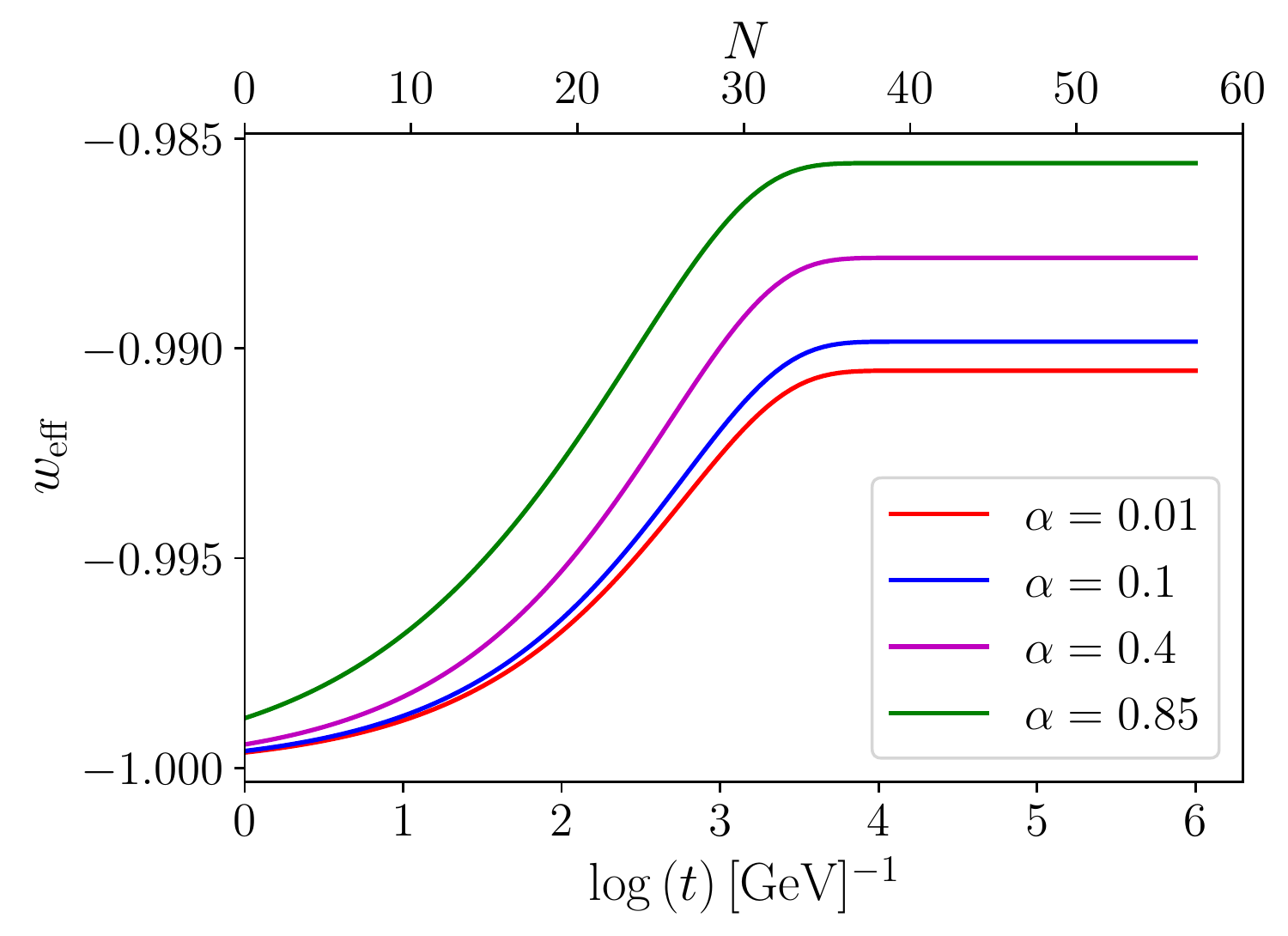}
  \includegraphics[width=0.5\textwidth]{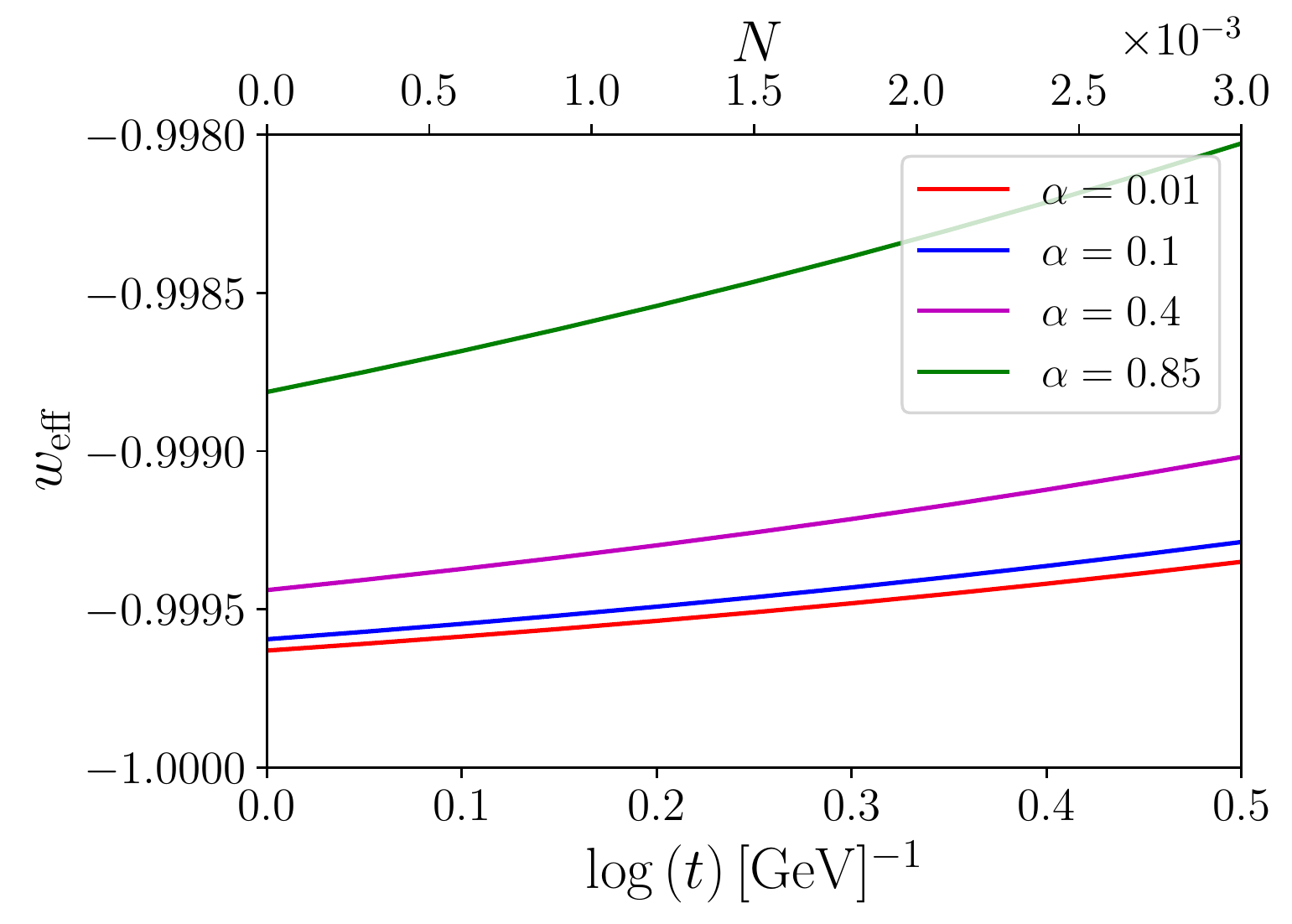}
  \includegraphics[width=0.5\textwidth]{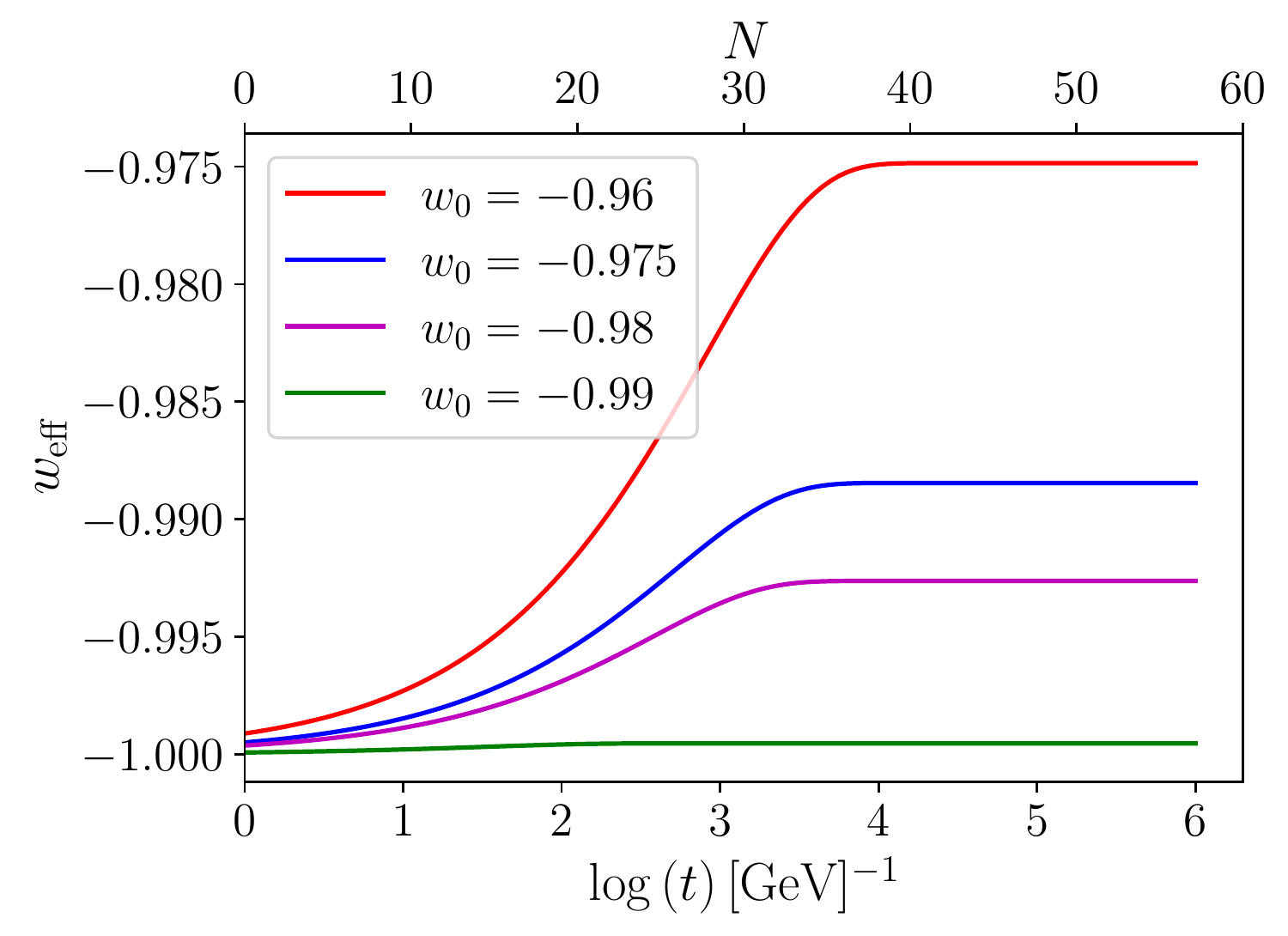}
  \includegraphics[width=0.49\textwidth]{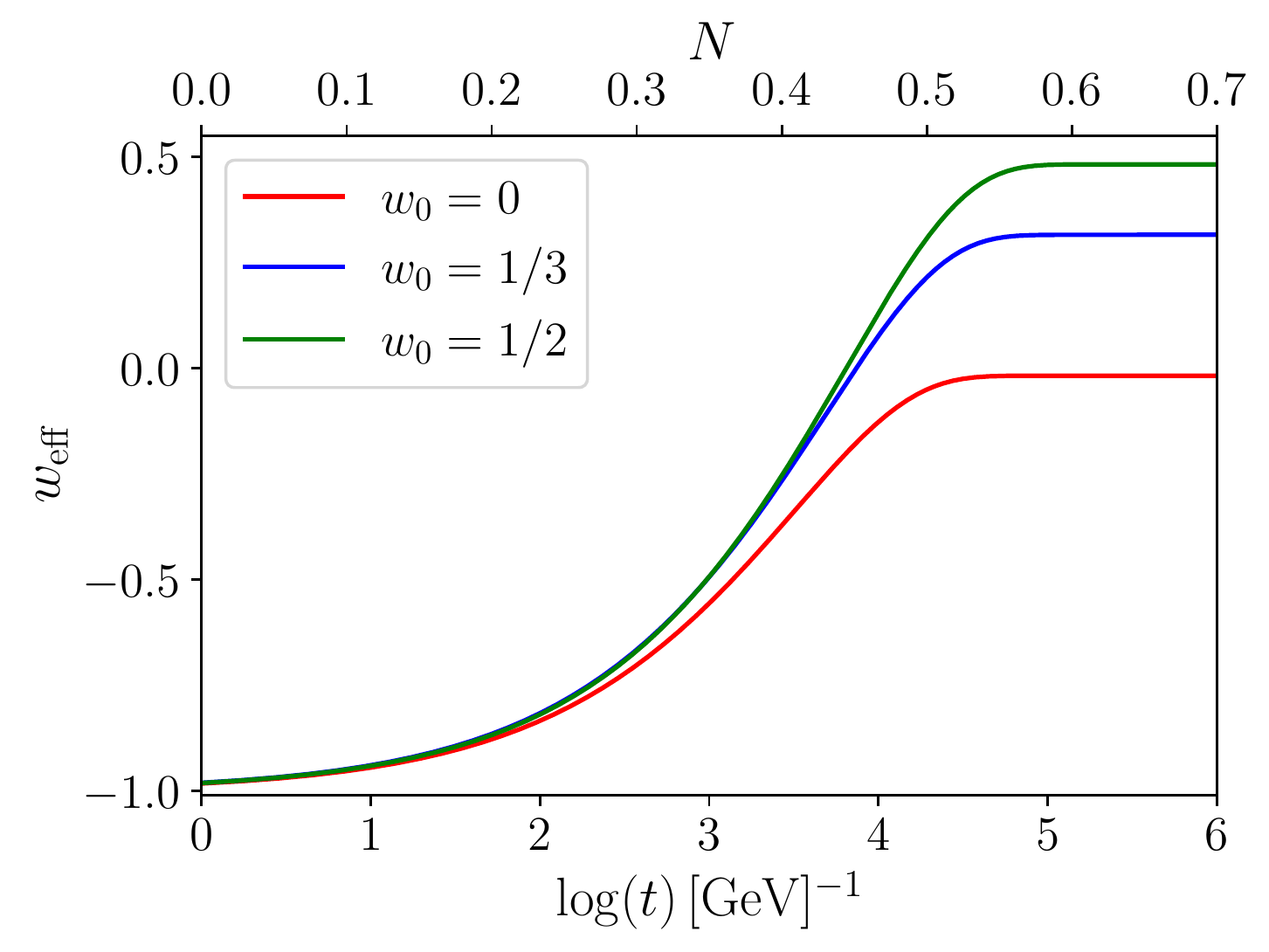} 
  \caption{\textit{Top row:} Evolution of $w_{\rm eff}$ for different
    values of $\bar{\zeta}$ for reference values of $\alpha =0.1$ and
    $w_{0} = -0.975$. The~right panel simply shows a magnification of the
    early evolution. \textit{Middle row:} the same as above but for
    different values of $\alpha$ while keeping fixed $w_0 = -0.975$ and
    $\bar{\zeta} = 0.01$. \textit{Bottom row:} the same as above but for
    different values of $w_0$ while keeping fixed $\bar{\zeta} = 0.01$
    and $\alpha=0.3$. In~all panels the top horizontal axes report
      time in terms of the number of e-foldings $N$.}
  \label{Fig-3}
\end{figure}

Figure~\ref{Fig-4} shows the evolution of $H(t)$ under the quasi
de-Sitter condition that $-\dot{H}\ll H^2$ for $\alpha =0.1$ and $w_0 =
-0.975$. Note that the Hubble function, which is initially set to take
the value at which inflation is believed to have taken place, \ie
$H=10^{16}\, {\rm GeV} \approx 10^{-3}\,M_{\rm{pl}}$, remains essentially
constant, but then, it starts to slowly fall off. During this period, as
shown by Figure~\ref{Fig-4}, the first Hubble slow-roll parameter remains
small and satisfies the condition $\epsilon_{\rm {H}}\ll1$. The behaviour
of $w_{\rm eff}$ is shown in the various panels of Figure~\ref{Fig-3} for
different values of the parameters $\bar{\zeta}$, $\alpha$, and $w_0$.

\unskip

In essence, all variations of the parameters show that, through the
time-evolving bulk viscosity pressure $\Pi(t)$, the effective
EOS parameter $w_\mathrm{eff}$ evolves with time and, finally, attains a nearly
constant value at a late time value during the quasi de-Sitter
inflationary period. More specifically, the effect of varying
$\bar{\zeta}$ on the effective EOS parameter is shown in the top row of
Figure~\ref{Fig-3}, where it is found that $w_\mathrm{eff}$ departs from the
exact de-Sitter value as $\bar{\zeta}$ is slowly decreased from its
maximum value $\bar{\zeta}_{\rm dS}$ ($\bar{\zeta}_{\rm dS}=0.0214$ for
$\alpha=0.1$ and $w_0=-0.975$). As a result, the more the magnitude of
$\bar{\zeta}$ is decreased, the greater the departure from the exact
de-Sitter evolution. Similarly, the influence of the variation of
$\alpha$ on $w_\mathrm{{eff}}$ is presented in the middle row of
Figure~\ref{Fig-3}, where it can be observed that, as $\alpha$ is increased,
$w_\mathrm{{eff}}$ deviates swiftly from the de-Sitter phase, implying
that an increase of the non-equilibrium characteristic timescale
$\tau_{*}$ leads to an increasing large departure from the exact
de-Sitter state. Stated differently, if non-equilibrium effects are present
for a sufficiently long time, the universe cannot remain close to an exact
de-Sitter state. Finally, the bottom row of Figure~\ref{Fig-3} reports the
behaviour of $w_\mathrm{eff}$ due to the variations of the parameter $w_0$
while keeping $\alpha$ and $\bar{\zeta}$ fixed. Since the value $w_0= -1$
is not allowed, decreasing $|w_0|$ leads to departures of $w_{\rm eff}$
from the exact de-Sitter phase for finite and positive values of
$\bar{\zeta}$ and $\alpha$.

\begin{figure}[H]
  \includegraphics[width=0.65\textwidth]{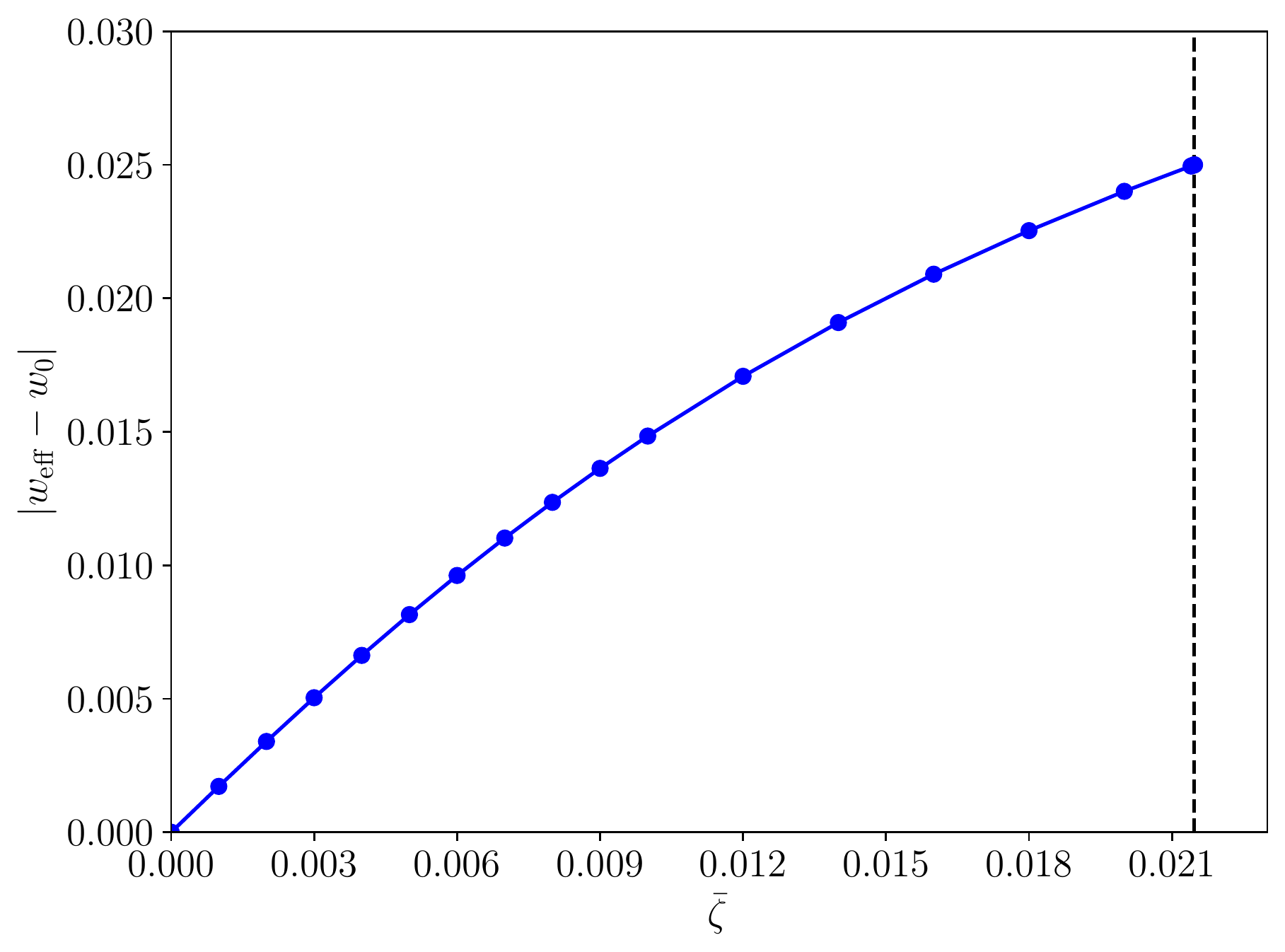}
  \caption{Variation of the absolute magnitude of the bulk-viscosity
    pressure $|w_{\rm{eff}}-w_0|$ as a function of the bulk-viscosity
    coefficient $\bar{\zeta}$ for reference values of $\alpha = 0.1$ and
    $w_0 = -0.975$; the vertical dotted line marks the maximum value
    $\bar{\zeta}_{\rm{dS}}= 0.02147$.}
  \label{weff-w0}
\end{figure}

It is worth noting that the profile of $w_{\rm {eff}}$ is a feature that
is purely a consequence of bulk viscous effects subject to quasi
de-Sitter conditions. In this regard, an insightful example is the
evolution when $w_0 =0$ (see the bottom-right panel of Figure~\ref{Fig-3}). In
this case, in fact, the evolution of $w_{\rm{eff}}$ depends only on
$\bar{\zeta}$ and $\alpha$. The effective EOS parameter exhibits a
departure from the exact de-Sitter phase, which is qualitatively very
similar to the cases when $w_0 \neq 0$. Furthermore, the late-time value of
$w_{\rm {eff}}$ increases systematically when decreasing $\bar{\zeta}$, and
it increases for larger values of $\alpha$ and $w_0$. Finally, we note
that, during the quasi de-Sitter expansion, the absolute magnitude of the
bulk viscosity pressure can be estimated using the quantity
$|w_{\rm{eff}}-w_0|$, which is reported in Figure~\ref{weff-w0}. As is in part
to be expected, the bulk viscosity pressure decreases as $\bar{\zeta}$
decreases from $\bar{\zeta}_{\rm{dS}}$ and becomes zero in the limit in
which $\bar{\zeta} \rightarrow 0$. All in all, the results presented in
Figure~\ref{Fig-3} clearly indicate that the effective EOS parameter
evolves following a unique functional behaviour, namely, from an almost
constant and small initial value to an almost constant and large final
value. However, the details on the transition between these two states is
a function of the three parameters involved in the evolution:
$\bar{\zeta}, \alpha$, and $w_0$.

\unskip

\section{Results and~Interpretation}
\label{sec-4}

In what follows, we summarise the bulk of the results obtained in our
numerical investigation and provide a global interpretation of the
various behaviours encountered. We start by considering what are the
ranges in which the parameters of our approach, namely $w_0$,
$\bar{\zeta}$, and $\alpha$, can vary.

\subsection{Admissible Regions of the~Parameters}

We first note that we have restricted our study to the values of the
EOS parameter such that $-1 \leq w_0 \leq 1$, so that we do not consider
the case $w_0<-1$ for which the scenario of phantom inflation is often
considered \cite{Capozziello:2005tf, Saridakis:2008fy,
  Khurshudyan:2016zse}. Clearly, using Equation~(\ref{eq-q:1}) with $w_0= -1$
yields $\dot{H}=0$, that is the standard and non-viscous \textit{exact}
de-Sitter universe (note that, in this case, $\bar{\zeta}_{\rm dS}=0$, as
  expressed by Equation~(\ref{eq-14})). On the other hand, when $w_0=1$,
Equation~(\ref{eq-q:1}) gives rise to $\dot{H} \neq 0$, so a
\textit{quasi} de-Sitter phase can in principle exist in this limit
provided the conditions $\dot{H}<0$ and $-{\dot{H}}/{H^2}\ll1$ are
valid. What remains problematic, however, is that, in this case, the
relaxation timescale $\tau_{_{\Pi}}$ diverges; for this reason, we will
not consider the value $w_0 =1$ as admissible in our analysis. Another
potentially problematic value for the EOS parameter is $w_0=1/2$, which
results in a diverging $\bar{\zeta}_{\rm{dS}}$
\cite{maartens1997nonlinear}; however, so long as the quasi de-Sitter
conditions remain in place, \ie $\dot{H}<0$ and $-{\dot{H}}/{H^2}\ll1$,
the issue of the divergence of $\bar{\zeta}_{\rm{dS}}$ at $w_0={1}/{2}$ does
not arise. This is shown in the bottom row of Figure~\ref{Fig-3}, where we
observe that $w_\mathrm{{eff}}$ is finite and well-behaved over the
entire interval $-1<w_0 < 1$ for different magnitudes of $\bar{\zeta}$.

Let us now consider the set of possible solutions starting from the
inviscid limit ($\bar{\zeta}=0$), which leads to $w_\mathrm{{eff}}= w_0$
\ie a constant effective EOS parameter. Under these conditions,
Equation~(\ref{eq-q:2}) reduces to
\begin{equation}
    \frac{3\sqrt{3}}{2}(w_0^2-1)\left(1+w_0\right)\left(
    \frac{2}{3}\frac{\dot{H}}{H^2}+1+w_0 \right) =0\,,
\end{equation}
whose possible solutions are:
\begin{itemize}
\item[\textit{(a)}] $w_0 = 1$: this is the perfect-fluid cosmological
  solution with an ultra-stiff EOS~\cite{Rezzolla_book:2013}.

\item[\textit{(b)}] $w_0 = -1$: this corresponds to the exact de-Sitter
  solution.

\item[\textit{(c)}] $\displaystyle\frac{2}{3}\frac{\dot{H}}{H^2}+1+w_0
  =0$: the corresponding solution is given by
\begin{equation}
    H(t) = \frac{2}{3 t (w_0+1)-2 c_1} \,,
    \label{A}
  \end{equation}
where the integration constant may be chosen to be $c_1=0$ or $c_1<0$ to
keep $H(t)$ finite. More specifically, if~$c_1=0$, the~scale-factor
is then given by
\begin{equation}
    a(t)\propto [3 t(1+w_0)]^{2/3 (w_0+1)}\,. \label{B}
\end{equation}

Equations~(\ref{A}) and (\ref{B}) thus provide straightforward
expressions for the evolution of the Hubble parameter and of the
scale-factor for either pressureless matter ($w_0 =0$) or for a
radiation-dominated ($w_0 = {1}/{3}$) universe. 
\end{itemize}

To determine the maximum allowed value of $\bar{\zeta}$,
Equation~(\ref{eq-q:2}) is re-expressed in terms of $\epsilon_{\rm H}$ (which
lies in the interval $0\leq \epsilon_{\rm H}\, \textless \,
1$). Substituting $w_\mathrm{eff}=-1+(2/3)\epsilon_{\rm H}$ in
Equation~(\ref{eq-q:2}) one arrives at 

\begingroup\makeatletter\def\f@size{9.5}\check@mathfonts
\def\maketag@@@#1{\hbox{\m@th\large\normalfont#1}}%
\begin{flalign} 
\frac{\dot{\epsilon}_{\rm H}}{3H(1-w_0^2)} +\frac{(1+w_0+\sqrt{3}\alpha
  \bar{\zeta})(-1+\frac{2}{3}\epsilon_{\rm
    H}-w_0)}{\sqrt{3}\bar{\zeta}[1+w_0+\alpha(-1+\frac{2}{3}\epsilon_{\rm
    H}-w_0)]}
+\frac{(1+w_{0})}{[1+w_0+\alpha(-1+\frac{2}{3}\epsilon_{\rm H}-w_0)]} -
\nonumber\\
\frac{(1-\frac{2}{3}\epsilon_{\rm H}+w_0)^2}{2(1+w_0)(1-w_0^2)}
=0\,. 
\end{flalign}\endgroup

The maximum value of the bulk viscosity $\bar{\zeta}$ required for
sustaining an accelerated expansion $\bar{\zeta}_{\rm{max}}$ is obtained
when considering $\epsilon_{\rm H}=0$ and $\dot{\epsilon_{\rm H}} \approx
0$ and is found to be
\begin{eqnarray}
   \bar{\zeta}_{\rm{max}}=\frac{2 \left(w_0^2-1\right)}{\sqrt{3} (1-\alpha)
     (2 w_0-1)} = \bar{\zeta}_{\rm dS}\,,
   \label{zeta-max}
 \end{eqnarray}
 which exactly matches with Equation~(\ref{eq-14}) and equals to the magnitude
 of the bulk viscosity that would give rise to an exact de-Sitter
 expansion. 

As a consistency check, we can set $\bar{\zeta}=\bar{\zeta}_{\rm{max}}$
in Equation~(\ref{eq-q:2}) to obtain
\vspace{-10pt}
\begin{adjustwidth}{-\extralength}{0cm}
\begin{equation}
    (w_{\rm{eff}}+1) \left[w_{\rm{eff}} (-\alpha -3 \alpha  w_0+w_0+1)+\alpha 
   w_{\rm{eff}}^2+(2 \alpha +1) w_0^2+2 w_0^3+(\alpha -3) w_0-2\right] =0\,,
\end{equation}
\end{adjustwidth}
which has three real roots one of which is $w_{\rm{eff}} =-1$. 
The other two roots are 
\begin{equation}
w_{\rm{eff}}=\frac{-\alpha +(w_0+1) \sqrt{\alpha ^2+\alpha (6-8 w_0)+1}-3
  \alpha w_0+w_0+1}{2 \alpha }\,,
\label{root1}
\end{equation}
and
\begin{equation}
w_{\rm{eff}}=\frac{\alpha +(w_0+1) \sqrt{\alpha ^2+\alpha (6-8 w_0)+1}+3
  \alpha w_0-w_0-1}{2 \alpha }\,,
\label{root2}
\end{equation}
and are not relevant here, since for Equation~(\ref{root1}) $w_{\rm{eff}}< -1$
in the range $0<\alpha<1$, while for Equation~(\ref{root2}) $w_{\rm{eff}}> 1$
in the range $-0.5<w_0<1$ and for all values of $\alpha$. As~a result,
$w_{\rm{eff}} =-1$ is the only possible solution of Equation~(\ref{eq-q:2})
during the exact de-Sitter phase when $\bar{\zeta} =
\bar{\zeta}_{\rm{max}}$.

\begin{figure*}
\includegraphics[width=0.49\textwidth]{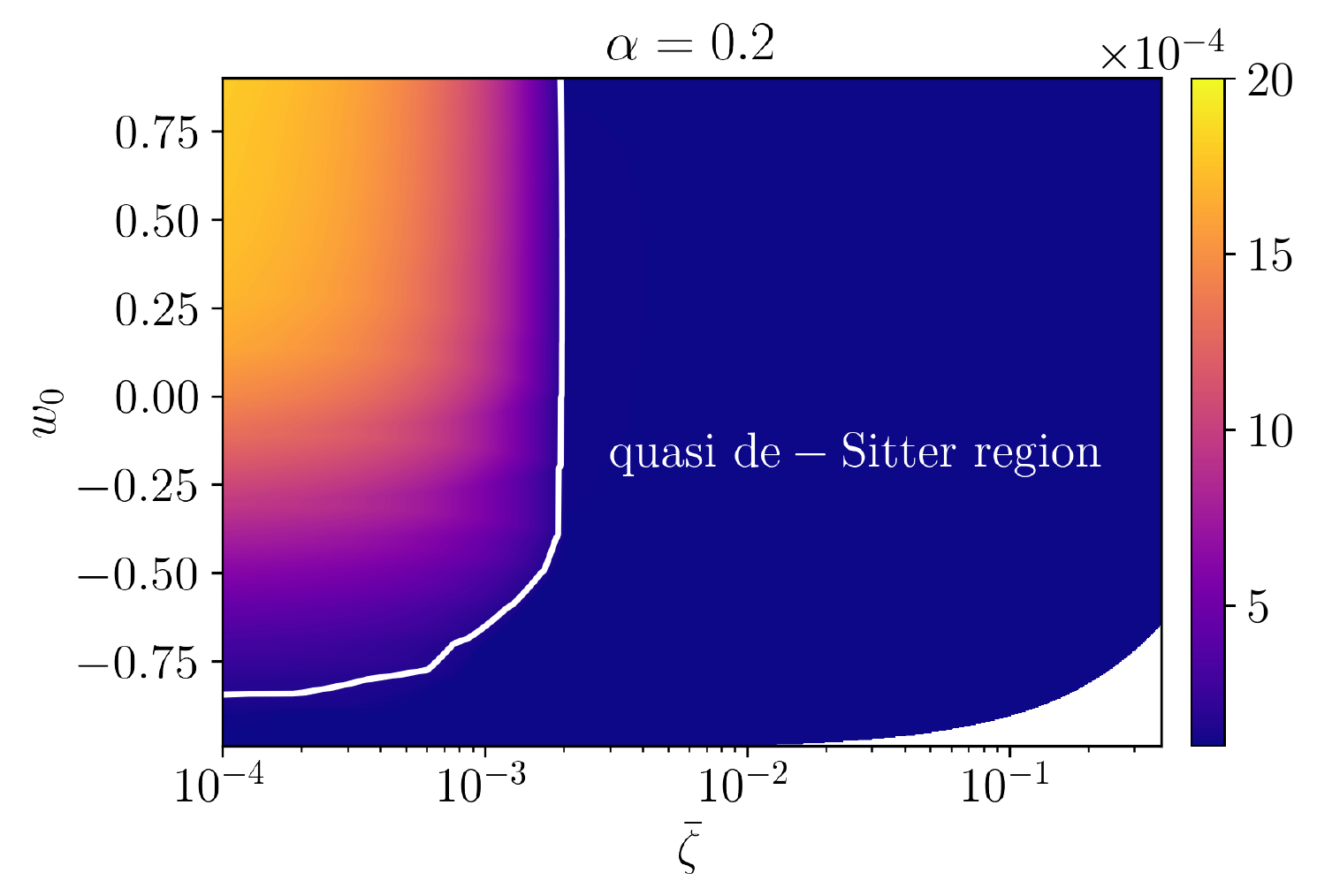} \label{a_c1}
\includegraphics[width=0.49\textwidth]{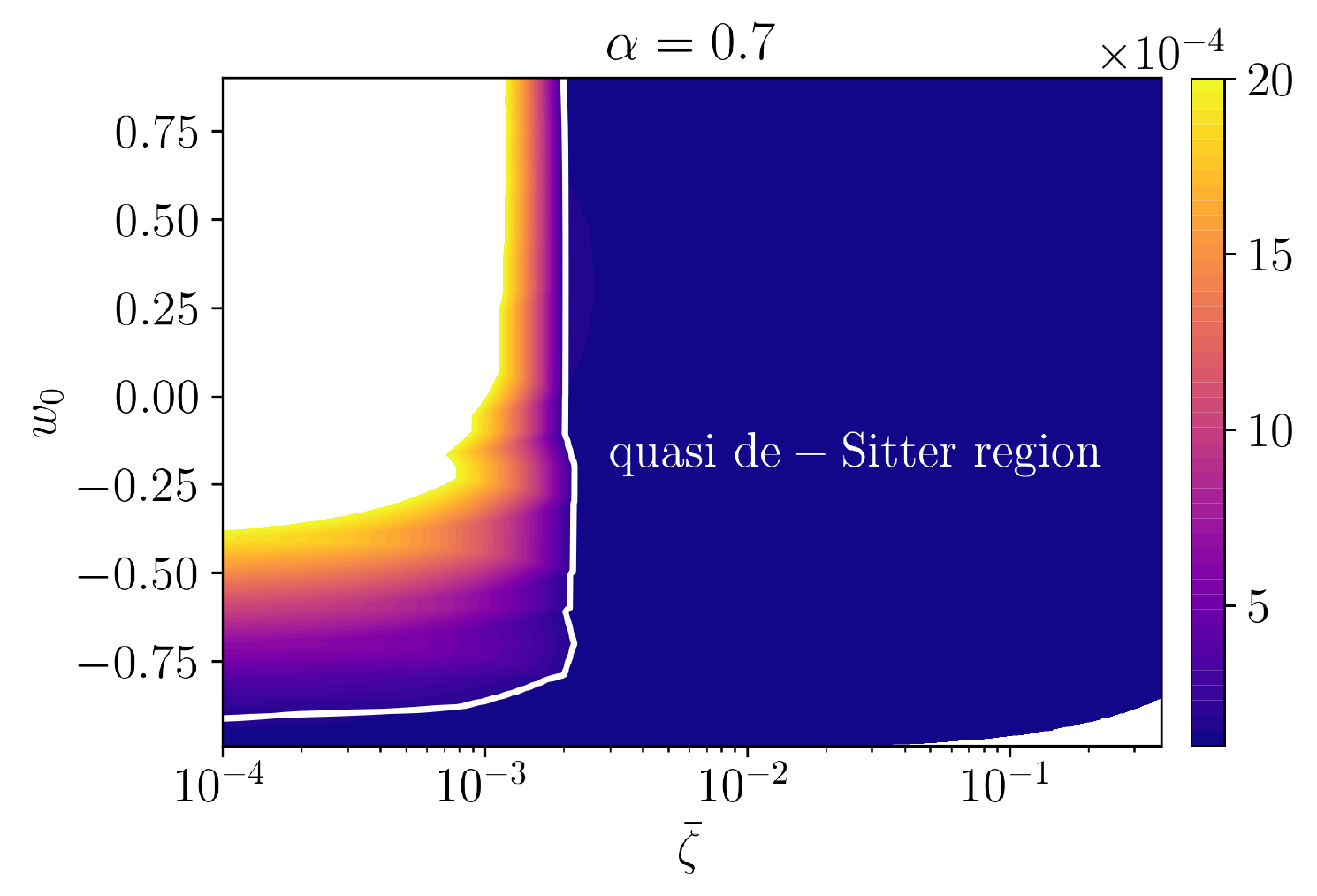} \label{a_c2}
\includegraphics[width=0.49\textwidth]{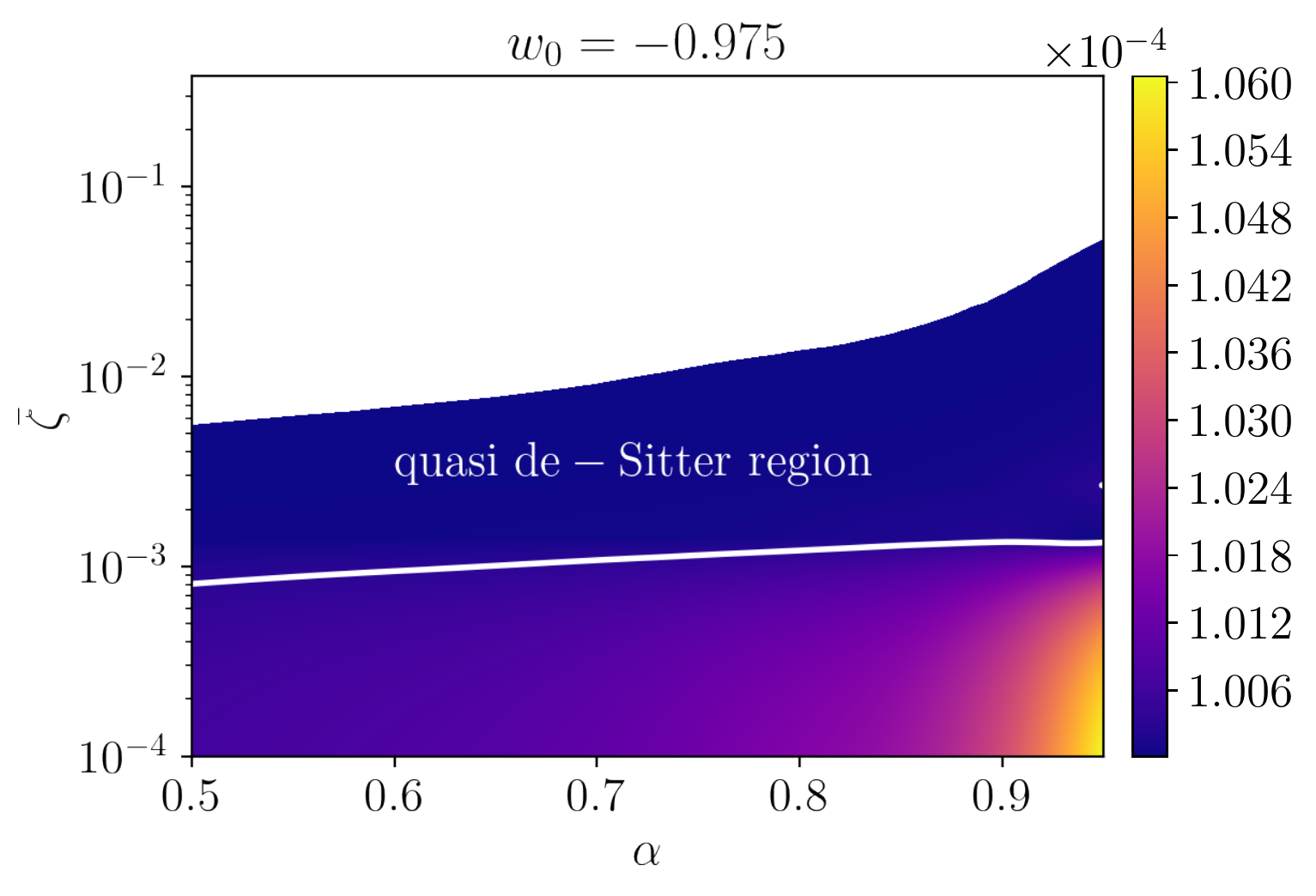} \label{w0_c}
\includegraphics[width=0.49\textwidth]{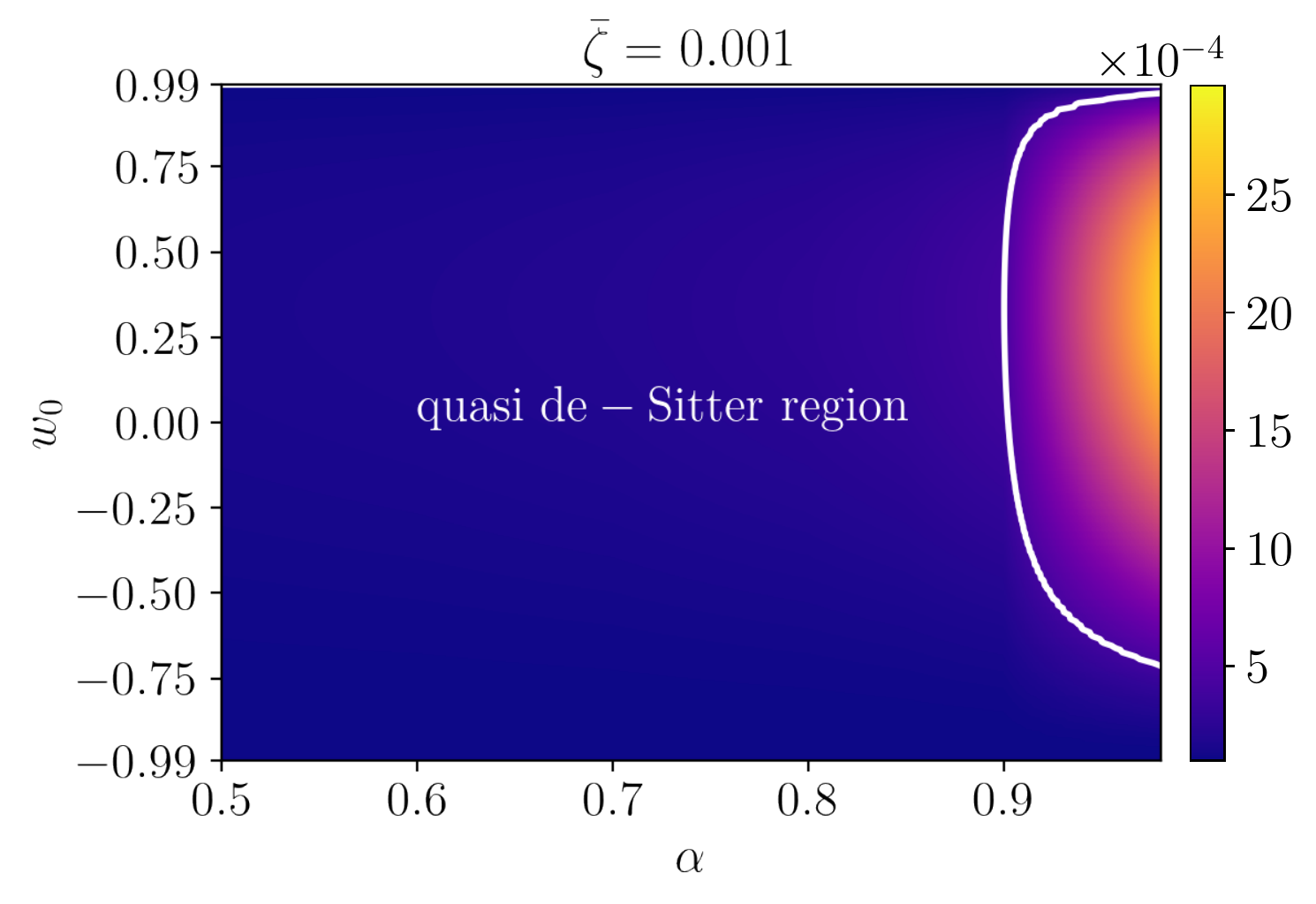} \label{zeta_c}
\caption{Colour maps of the measure of the deviation from the quasi
  de-Sitter condition via the quantity $-\dot{H}/{H^2}$ ($-\dot{H}/{H^2}
  \ll 1$ corresponding to a quasi de-Sitter expansion; the white line
  marks the critical value $-\dot{H}/H^2= 0.00016$. The maps are obtained
  by solving Equation~(\ref{eq-q:1}) for different magnitudes of the
  parameters $\bar{\zeta}$, $w_0$, and $\alpha$, with the first row
  referring to scenarios in which $\alpha=0.2$ (left) and $\alpha=0.7$
  (right). The bottom row is the same as the top one but when considering
  constant values of $w_0 =-0.975$ (left) and $\bar{\zeta} = 0.001$
  (right), respectively.}
  \label{fig:colormaps_hdot}
\end{figure*}

In order to represent the effects of the variations of $w_0$,
$\bar{\zeta}$ and $\alpha$ on the solution obtained during the quasi
de-Sitter inflationary period, we present in
Figure~\ref{fig:colormaps_hdot} with colour maps the values of
$-\dot{H}/H^2$ obtained when varying either $\bar{\zeta}$ or $\alpha$
while keeping one of these quantities fixed. In~each case, we
considered the admissible ranges of the parameters $\alpha, \bar{\zeta}$,
and $w_0$ and found that quasi de-Sitter solutions exist over the entire
interval $-1<w_0<1$ of $w_0$. However, the~top panels of
Figure~\ref{fig:colormaps_hdot} also show that significant deviations from
the quasi de-Sitter evolutions can take place if $\bar{\zeta}$ is kept
small. More specifically, while for large negative values of $w_0$ (\ie
$w_0 \lesssim -0.81$), a de-Sitter evolution is possible even for large
values of the bulk viscosity, as~the EOS parameter increases, a~substantial amount of bulk viscosity is necessary to ensure a de-Sitter
evolution. In~particular, setting as a critical value $-\dot{H}/H^2=
0.00016$, we found the a de-Sitter evolution is possible only for
$\bar{\zeta} \gtrsim 10^{-4}$ (see left panel of
Figure~\ref{fig:colormaps_hdot}). Interestingly, for~$w_0 \gtrsim -0.25$,
the amount of bulk viscosity needed to ensure a de-Sitter evolution is
essentially constant a corresponds to $\bar{\zeta} \sim 2\times
10^{-3}$. Stated differently, while a quasi de-Sitter evolution is
possible for any value of $w_0$, if~the bulk viscosity is large, the~latter is small, only a delicate balance is required between $w_0$ and
$\bar{\zeta}$. Note also that the considerations made above depend --
although not sensitively -- on the value of $\alpha$, as~can be
appreciated when comparing the left and right top panels of
Figure~\ref{fig:colormaps_hdot}. More specifically, the~regions of quasi
de-Sitter evolution tend to be slightly reduced as $\alpha$ is~increased.

The lower-left panel of Figure~\ref{fig:colormaps_hdot}, on the other hand,
shows that the impact of the $\alpha$ parameter is small so long as $w_0
\sim -1$ and the deviation from the quasi de-Sitter phase remains small
over the entire range of $\alpha$. For instance, a slight deviation from the
quasi de-Sitter phase takes place only for the
bulk viscosity coefficient around $\bar{\zeta} \lesssim 10^{-3}$, with
the deviation becoming comparatively larger as $\alpha$ becomes close to
unity. However, for bulk viscosities $\bar{\zeta} \gtrsim10^{-3}$, a
quasi de-Sitter phase always exists independently of the value of
$\alpha$. Finally, the lower-right panel of
Figure~\ref{fig:colormaps_hdot} reports the deviation from the exact de-Sitter phase
when considering a fixed bulk viscosity coefficient $\bar{\zeta}=0.001$
and varying the two coefficients $w_0$ and $\alpha$. In this case, it is
clear that the quasi de-Sitter expansion takes place for all values of $w_0$
so long as $\alpha \lesssim 0.9$. However, when $\alpha \gtrsim 0.9$,
deviations from the de-Sitter phase
 become large for most of the range of $w_0$ and,
in particular, if $w_0 \geq -0.75$.

Given that $w_{\rm{eff}}$ does not remain constant during the quasi
de-Sitter inflation, but increases till reaching a new stationary value
(see Figure~\ref{Fig-3}), it is reasonable to define a timescale $\tau$
that characterises how rapid the transition is from the exact de-Sitter
phase to the new quasi de-Sitter inflation. We use the behaviour shown
in Figure~\ref{Fig-3} to express the evolution of $w_\mathrm{eff}$ as
\begin{equation}
  \tilde{w}_\mathrm{eff}(x)=-1+\frac{A}{1+Be^{-x/\tau} }
  \label{f-eqn}\,,
\end{equation}
where $x:=\log_{10}t$ and the dimensionless constants $A, B$ depend on
the bulk viscosity. Using a best-fit approach and the numerical results
presented in Figure~\ref{Fig-3}, it is possible to compute both $\tau$ and
$A, B$. Unsurprisingly, and just like $w_{\mathrm{eff}}$, also $\tau$
depends on $\bar{\zeta}$, $\alpha$, and $w_0$, so that, for $\alpha$ and
$w_0$ fixed, $\tau$ simply decreases as $\bar{\zeta}$ is increased up to
$\bar{\zeta}_{\rm{dS}}$. Indeed, for $w_0 \rightarrow -1$, a considerable
time is needed for $w_\mathrm{eff}$ to deviate from the exact de-Sitter
phase (see the bottom-left panel in Figure \ref{Fig-3}). This behaviour of
$w_\mathrm{eff}$ is reflected in $\tau$ as well, which increases only
very slowly even if $\bar{\zeta} \ll \bar{\zeta}_{\rm{dS}}$.

\subsection{Estimation of the Model~Parameters}

We next turn our attention to estimating the allowed values of these
parameters by using constraints from the primordial power spectrum generated
by curvature perturbations during the inflationary period. Such an
analysis will not only enable us to estimate realistic values of the
bulk-viscosity coefficient and characteristic timescales, but~will also
allow us to examine the observational viability of the quasi de-Sitter model
of inflation supported by bulk viscosity. A~straightforward procedure in
this direction involves the evaluation of the inflationary variables, \eg
the spectral index of scalar curvature perturbations $n_s$ and the
tensor-to-scalar ratio $r$, in~terms of the model parameters
$\bar{\zeta}$, $\alpha$, and $w_0$, as well as~the comparison with the
experimentally observed values of $n_s$ and $r$. 
A possible way to do this is to construct the fluid
description of inflation in the context of the generalised causal theory
following the procedure suggested by Bamba and
Odintsov~\cite{bamba2016inflation}, who introduced an EOS which included
the bulk viscosity, but did not consider a causal theory of~hydrodynamics.

Typically, the~inflationary variables are expressed in terms of slow-roll
parameters which in turn are represented in terms of the potential of a
scalar field driving inflation and their respective derivatives. The~equivalence between the fluid description of inflation and the
scalar-field description of inflation can be established provided the EOS
of the bulk viscous fluid and that of the equation of the state followed
by the pressure and energy density of a scalar field are the
same~\cite{bamba2016inflation} ({A similar} 
 strategy can be
  found in~\cite{Zimdahl:1999tn} which addresses bulk viscosity driven
  inflation using MIS theory by taking into account of particle
  production.).  Hence, we will first elucidate the equivalence between
the two~descriptions.


We recall that the Friedmann equations when expressed in terms of the
e-folding number $N$ are given by
\begin{eqnarray}
3 (H(N))^2&=& e(N)\,,   \label{N-rho} \\
2 H(N)H'(N)+3 (H(N))^2&=&-p_{\mathrm{eff}}(N)\,, \label{N-p} 
\end{eqnarray}
where $N$ is defined as $N : = \int^{t_f}_{t_i} Hdt$, so that the Hubble
function becomes $H = H(N)$. Here, $t_{i}$ and $t_{f}$ are respectively
the times corresponding to the beginning and end of inflation, while the
index $'$ is used to indicate the derivative with respect to $N$. From~Equation~(\ref{N-rho}) and Equation~(\ref{N-p}), we then obtain
\begin{eqnarray}
\frac{1}{2}\left[e(N)-p_\mathrm{{eff}}(N)\right] =H'(N)H(N)+3(H(N))^2\,. \label{eq1}
\end{eqnarray}
{On} the other hand, in~the single scalar-field description of inflation,
the action is given by
\begin{eqnarray}
S := \int d^4 x \sqrt{-g} \left(\frac{R}{2}-\frac{1}{2} \partial_{\mu}\phi
\partial^{\mu}\phi-V(\phi)\right)\,,
\end{eqnarray}
where $\phi = \phi(t)$ is also known as the inflaton field and $V(\phi)$
is the inflaton potential and $\sqrt{-g}$ is the determinant of the metric. The~components of the Einstein equations are
\begin{eqnarray}
3H^2 &=& \frac{1}{2}\dot{\phi}^2 +V(\phi)= e_{\phi} \,,\label{scalar-rho}
\\ -(2\dot{H}+3H^2)&=& \frac{1}{2}\dot{\phi}^2 -V(\phi)=
p_{\phi}\,, \label{scalar-p}
\end{eqnarray}
where $e_{\phi}$ and $p_{\phi}$ are the energy density and the pressure
of the scalar field, respectively. Combining these two equations one
obtains that
\begin{equation}
    \dot{\phi}^2 = -2\dot{H}\,.
\end{equation}

The equivalence between two descriptions of inflation is obtained once
the scalar field $\phi$ and the cosmological time $t$ are both rescaled
by an auxiliary scalar field $\Phi$, thus implying $\phi=\phi(\Phi)$ and
$t=t(\Phi)$, and~when $\Phi$ identified with the e-folding number
$N$. Under~these conditions, Equation~(\ref{scalar-rho}) and
Equation~(\ref{scalar-p}) can be rewritten as,
\begin{eqnarray}
3 H^2(N)&=& \frac{1}{2} H^2(N) \left(\frac{d \phi}{d\Phi}\right)^2
\Bigg|_{\Phi=N} +V(\phi(\Phi))|_{\Phi=N}\,, \\
-2 H(N)H'(N)&=&\frac{1}{2}
H^2(N) \left(\frac{d \phi}{d\Phi}\right)^2\Bigg|_{\Phi=N}
-V(\phi(\Phi))|_{\Phi=N}\,. \label{phi-2}
\end{eqnarray}

Using Equations~(\ref{scalar-rho})--(\ref{phi-2}), the~scalar-field potential
is given by 
\begin{eqnarray}
\label{Potential} V(N)&=&H'(N)H(N)+3H^2(N) = \frac{1}{2}(e_{\phi}-p_{\phi})\,.
\end{eqnarray}
{Since} the bulk viscous fluid and the scalar field possess the same
EOS, together with Equation~(\ref{eq1}) the potential can be
expressed as
\begin{eqnarray}
V(N)&=&H'(N)H(N)+3H^2(N) =\frac{1}{2}\left[e(N)-p_{\mathrm {eff}}(N)\right]\,.
\end{eqnarray}
and Equation~(\ref{eq-q:2}) can be re-written in the following way in terms of
$N$, namely,
\begin{flalign}
\tau_{_{\Pi}} \frac{d \Pi(N)}{dN} \left(1+k^2\frac{ \tau_{_{\Pi}}}{\zeta(N)} \Pi(N)
\right)+\frac{\Pi(N)}{H(N)}\left(1+3k^2H(N)\tau_{_{\Pi}}\right) \nonumber
\\ =-3\zeta(N)-\frac{1}{2}\tau_{_{\Pi}}\Pi(N)\left[3+\frac{1}{\tau_{_{\Pi}}}\frac{d
    \tau_{_{\Pi}}}{dN} -\frac{1}{\zeta}\frac{d
    \zeta}{dN}-\frac{1}{T}\frac{dT}{dN}\right]\left(1+k^2\frac{
  \tau_{_{\Pi}}}{\zeta(N)} \Pi(N) \right)\,, \label{master-eqn-N}
\end{flalign}
where the bulk viscosity is expressed as
\begin{eqnarray}
\Pi(N)&=&-2H'(N)H(N)+3(1+w_0)H^2(N)\,.
\end{eqnarray}
Similarly, the~relaxation time coefficient, the~temperature and the bulk
viscosity can also be expressed in terms of the e-folding number. Using
the potential, it is then possible to determine the slow-roll parameters
$\epsilon_{\rm V}(N)$ and $\eta_{\rm V}(N)$ purely in terms of $H(N)$,
$H'(N)$ and $H''(N)$ as~\cite{Bamba:2014daa}
\begin{flalign}
\epsilon_{\rm V}(N)=-\frac{H}{4H'}\left[\frac{6{H'}/{H} +
    {H''}/{H} +
    \left({H'}/{H}\right)^2}{3+{H'}/{H}}\right]^2\,,
\end{flalign}
and 
\vspace{-10pt}
\begin{adjustwidth}{-\extralength}{0cm}
\centering 
\begin{flalign}
\eta_{\rm V}(N)=-\frac{1}{2} \left[\frac{9{H'}/{H} +
    3{H''}/{H} + \frac{1}{2}\left({H'}/{H}\right)^2 -
    \frac{1}{2}\left({H''}/{H'}\right)^2 +3\left({H''}/{H'}
    \right)+ {H'''}/{H'}\,}{\left(3+{H'}/{H}\right)}\right]\,.
\end{flalign}
\end{adjustwidth}
{Typically}, the~energy scale of inflation and the e-folding number are
respectively taken to be $H_{in} \simeq 10^{-3}\, M_{\rm{Pl}}$, where
$M_{\rm{Pl}}$ is the Planck mass, and~$N\simeq 50$. The~energy scale of
inflation becomes $H_{in}= 10^{-3}$. Solving Equation~(\ref{master-eqn-N})
numerically under the condition that $-H'(N)\ll H(N)$ (or, equivalently,
that $\dot{\phi}^2 \ll V$ in a scalar-field description), we have found
that $\epsilon_{\rm V}(N)$ and $\eta_{\rm V}(N)$ remain constant and
smaller than unity when $N$ is taken to vary in the interval $N\in [0,
  50]$ and $\alpha, w_0$ and $\bar{\zeta}$ are varied appropriately (see the 
discussion in Sec.~\ref{sec:cwo}). This suggests that slow-roll
conditions are indeed obeyed during the quasi de-Sitter inflationary phase
described by a generalised causal theory. Consequently, the~inflationary
observables $n_s$ and $r$ that characterise the perturbation spectra can
then be related to the potential slow-roll parameters $\epsilon_{\rm
  V}(N)$ and $\eta_{\rm V}(N)$ as follows (see, \eg~\cite{Riotto:2002yw})
\begin{eqnarray}
n_s&=&1-6\epsilon_{\rm V}+2\eta_{\rm V} \,,\label{ns}\\
r&=&16 \epsilon_{\rm V}  \label{r}\,.
\end{eqnarray}
\textcolor{black}{{We} note that, while for exact de-Sitter expansion, the~de-Sitter symmetries predict $n_s=1$
and $r=0$}, observations from the Cosmic Microwave Background (CMB) strongly
suggest a small but non-zero deviation of $n_s$, thus suggesting a scale
dependence of the density fluctuations and indicating a nearly de-Sitter
expansion, as well as~a small, but nonzero value for $r$.

\begin{figure}[H]
\includegraphics[width=0.65\textwidth]{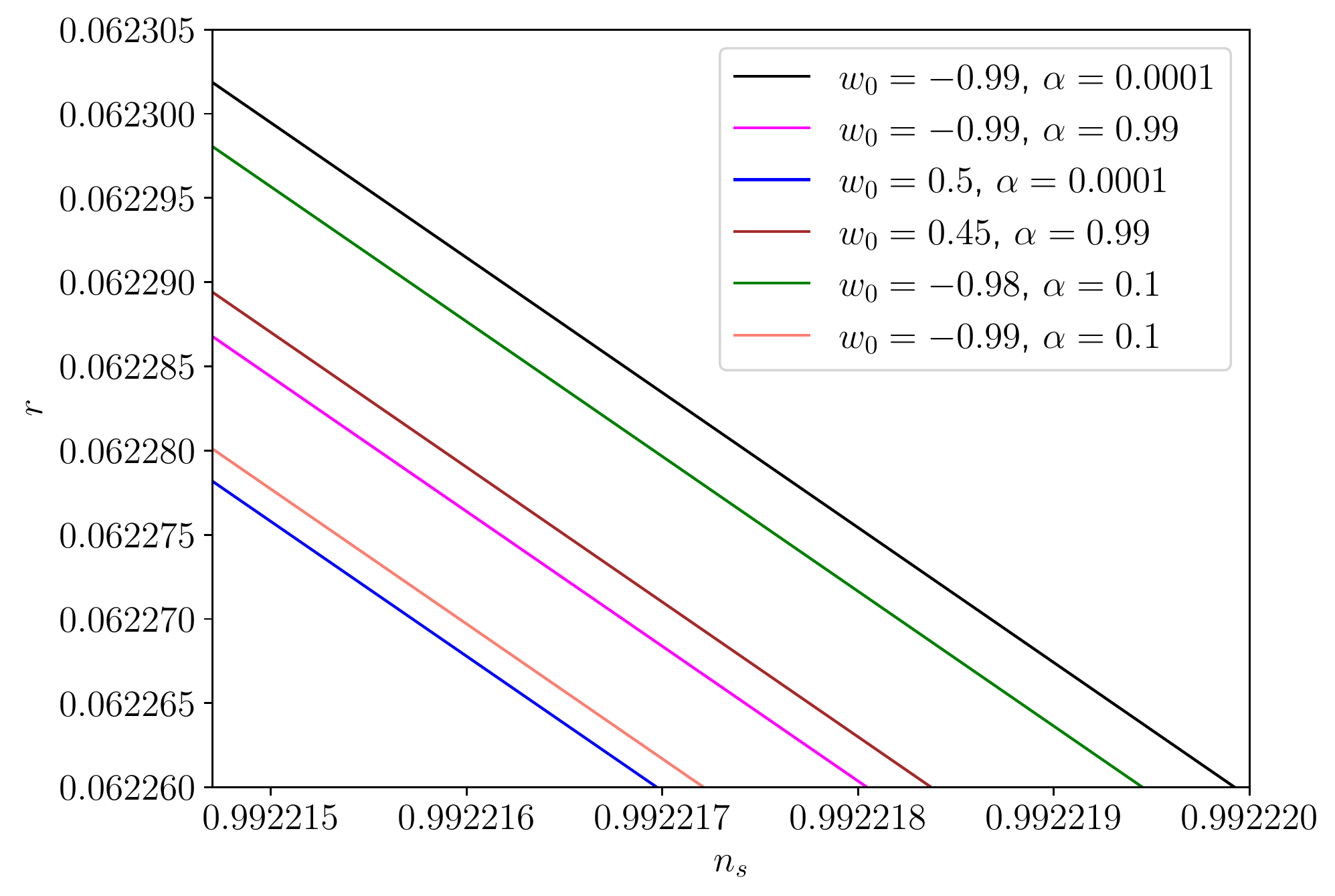} 
\caption{Behaviour of the spectral index $n_s$ and of the
  tensor-to-scalar ratio $r$ as predicted by our model. These quantities
  are evaluated using Equations~(\ref{ns})--(\ref{r}) for different values of
  $w_0$, $\alpha$ and $\bar{\zeta}$ when they are varied within the
  corresponding allowed ranges.}
\label{zoom}
\end{figure}

The behaviour of $n_s$ and $r$ as predicted by our model is illustrated
in Figure {\ref{zoom}}, where we evaluated these quantities using
Eqs. (\ref{ns}) and (\ref{r}) for different values of $w_0$, $\alpha$, and
$\bar{\zeta}$ when they are varied within the corresponding allowed
ranges. Note that the results show an essentially linear behaviour and
that each point of a specific line corresponds to a unique value of
$\bar{\zeta}$ for fixed values of $\alpha$ and $w_0$, increasing from
left to right as $n_s$ is increased. Note also that, while each individual
line represents a unique solution for fixed values of $\alpha$ and $w_0$,
the actual variations in the spectral index are rather small, and this
explains the expanded scale shown in Figure~\ref{zoom}. In turn, this
implies that the effective space of allowed values for $n_s$ and $r$ is
essentially one-dimensional in our model and that, in the limit $n_s \to
1$, the tensor-to-scalar ratio is also vanishingly small, \ie $r\to 0$.

\subsection{Comparison with~Observations}
\label{sec:cwo}

\begin{figure}[H]
  \includegraphics[width=0.32\textwidth]{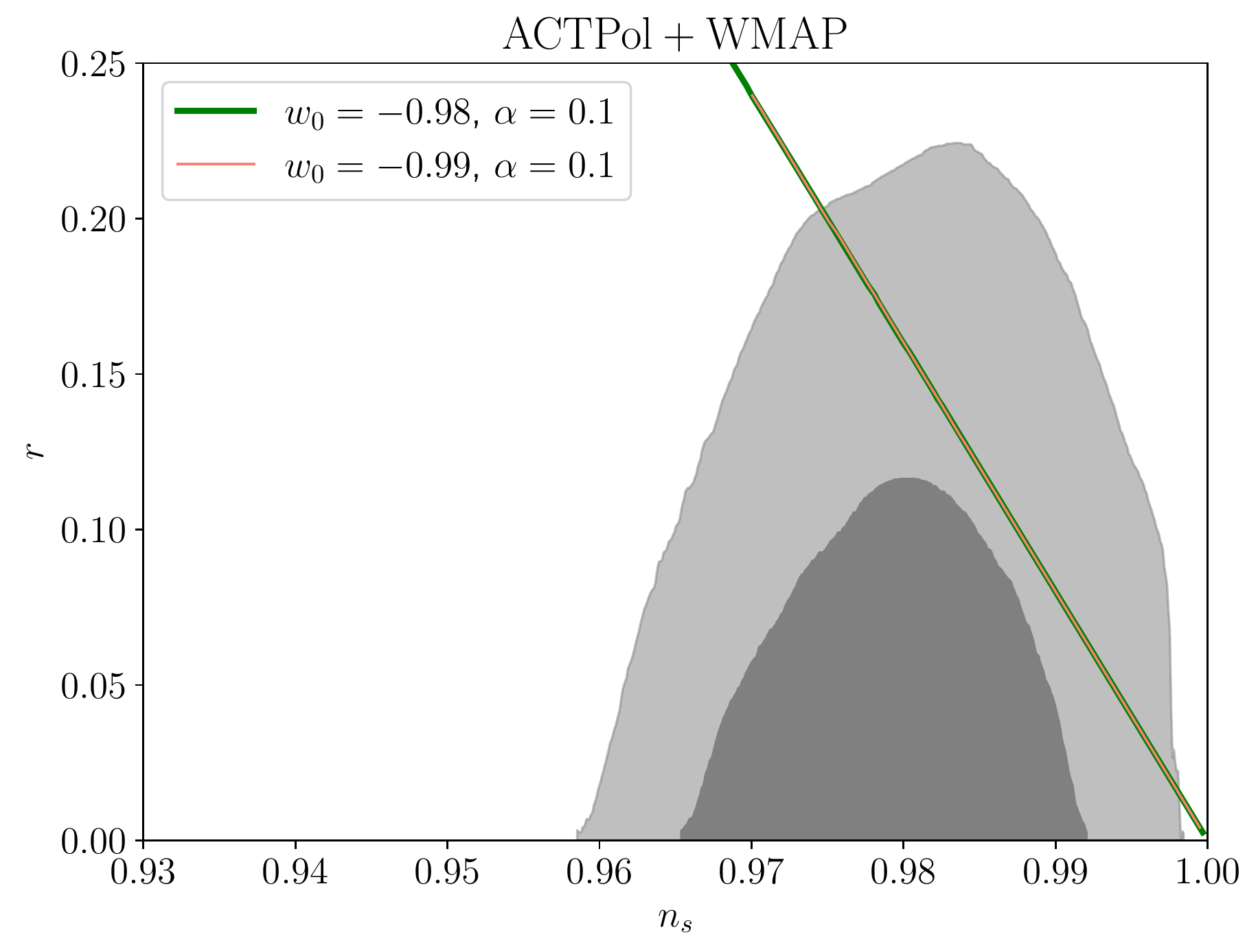}  \label{ACT}
  \includegraphics[width=0.32\textwidth]{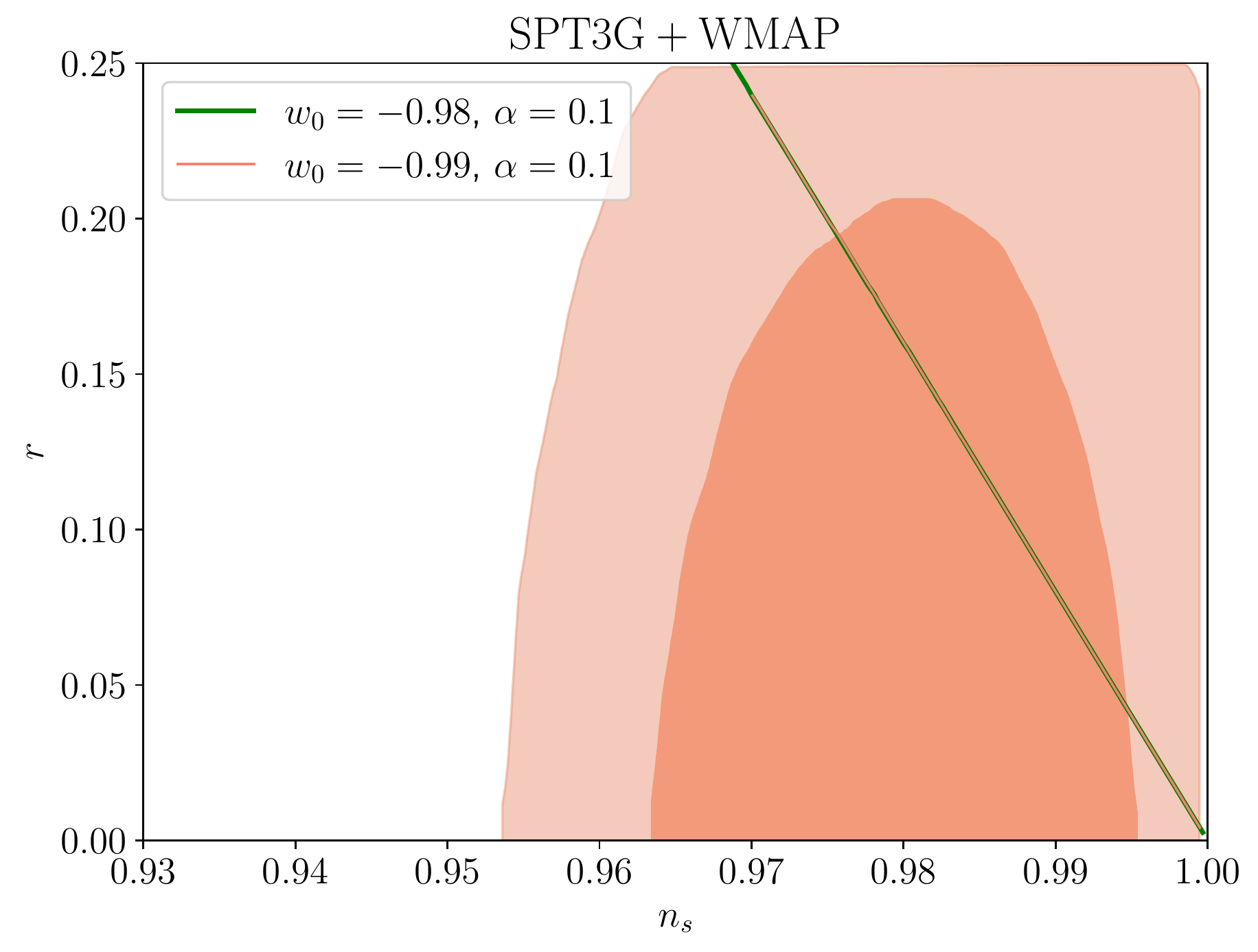}  \label{SPT} 
  \includegraphics[width=0.32\textwidth]{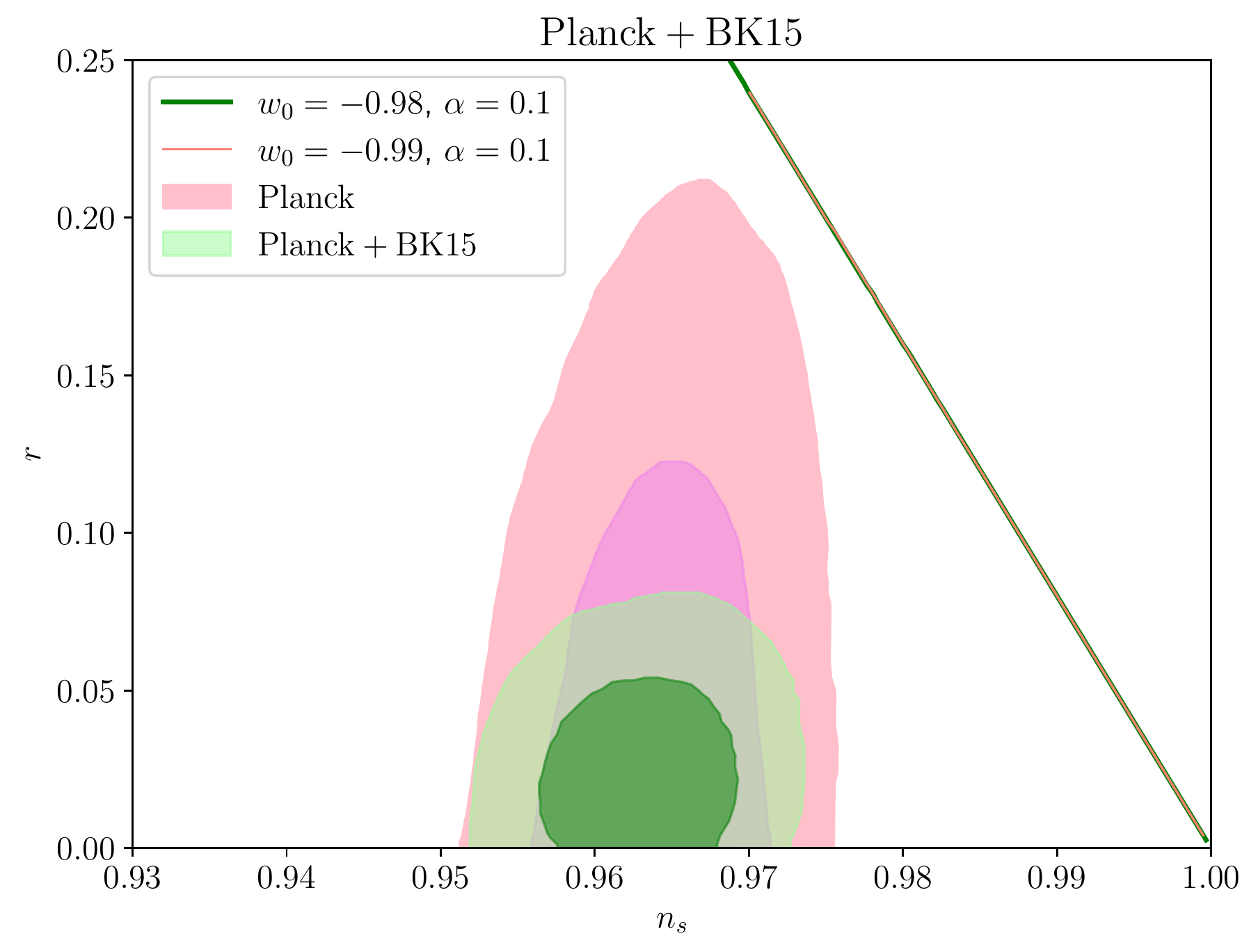}  \label{planck}
\caption{Comparison of bulk-viscous model of inflation (straight lines)
  with the marginalised contours of Planck$+$BK15, ACTPol$+$WMAP and
  SPT3G$+$WMAP results. The~contours are obtained for $n_s$, $r$ using
  marginalised joint 68$\%$ and 95$\%$ confidence-level regions from the
  observational data.}
\label{ns-rPlot}
\end{figure}

In order to examine the compatibility of our model with current
observational data, we considered three datasets obtained on observations
from the Planck satellite and from other, non-Planck-based
observations. More specifically, we considered the observational
constraints on $n_s$ and $r$ based on the Planck2018$+$ Keck-Array BK15
(``Planck$+$BK15'') data sets \cite{Planck:2019nip,
  Planck:2018vyg, Planck:2018nkj, BICEP2:2018kqh}, from the Atacama
Cosmology Telescope DR4 likelihood combined with the WMAP satellite data
set (``ACTPol$+$WMAP'') \cite{2013ApJS..208...19H,ACT:2020gnv},
and from the South-Pole Telescope polarization measurements (``SPT3G$+$WMAP'') \cite{2013ApJS..208...19H,SPT-3G:2021eoc}. The
comparison of our model with these three observational datasets was
carried out by considering two reference viscous cosmology datasets,
namely $w_0=-0.98, \alpha =0.1$ and $w_0=-0.99, \alpha =0.1$, which are shown in the three panels of Figure~\ref{ns-rPlot} (given the small
variance, they appear as solid straight lines in the panels). Also
reported are the constraints from the various datasets, namely
ACTPol$+$WMAP, SPT3G$+$WMAP, and Planck$+$BK15, from left to right. The
comparison in Figure~\ref{ns-rPlot} shows that the proposed model of
inflation is clearly compatible with both the ACTPol$+$WMAP and the
SPT3G$+$WMAP results, since the straight lines generated by the two
reference viscous cosmology datasets pass through the shaded regions in
the $(n_s, r)$ plane.

\vspace{-10pt}

Furthermore, we note that when $w_0 \rightarrow -1$, it is possible (and
straightforward) to select values of $\bar{\zeta}$ and $\alpha$ within
their allowed ranges of variation such that $n_s$ and $r$ match the
observational constraints. On~the other hand, when $w_0 > 0$, the~bulk-viscosity coefficient needed is very large (typically ${\mathcal
  O}(1)-{\mathcal O}(10)$) in order to produce compatible values of $n_s$
and $r$ ({When} considering $w_0 > 0.5$, the~values of $n_s$ and
$r$ obtained easily fall out of the range allowed by observations. For~example, for~$w_0=0.52$, we obtain \mbox{$n_s = 0.9061$} and \mbox{$r =0.7455$.)}. Note
also that since Figure~\ref{ns-rPlot} shows that the observations from
ACTPol$+$WMAP predict a lower value of $r$ when compared to the the
SPT3G$+$WMAP data, the~magnitude of the bulk-viscosity coefficient
required to produce compatible values of $n_s$ and $r$ will be different
for the observational datasets. For~instance, with~$w_0=-0.98, \alpha
=0.1$, a~bulk viscosity coefficient $\bar{\zeta} \sim 7\times 10^{-3}$
($\bar{\zeta} \sim 6\times 10^{-3}$) is necessary to produce values of
$n_s$ and $r$ compatible with the constraints from the ACTPol$+$WMAP
(SPT3G$+$WMAP) observations.

\textls[-35]{Finally, it should be remarked that, although our model is supported by
the ACTPol$+$WMAP and SPT3G$+$WMAP} results, it is incompatible with the
current Planck$+$BK15 datasets (third panel of Figure \ref{ns-rPlot}) for
all the values of $\alpha$ and $w_0$ that are allowed. In particular, the
Planck$+$BK15 observations seem to predict values that are
systematically smaller than that obtained in our model for all of the
values that the spectral index is allowed to take.

It is worthwhile to note that our results are indeed in agreement with
those presented in Refs.~\cite{Forconi:2021que,Giare:2022rvg}, where a
higher value of $n_s$ was also found and, thus, in accord with the
constraints from ACTPol$+$WMAP and SPT3G$+$WMAP. In this sense, unless
the differences are due to yet-undetected systematic differences among
the observations, the bulk viscosity model of inflation suggests a
possible and slight modification of the $\Lambda$CDM model
scenario~\cite{Handley:2020hdp}. On the other hand, the discrepancy in
the value of $n_s$ may also be related to the simplified phenomenological
model adopted in this work for incorporating out-of-equilibrium effects
during the inflationary period of the early universe. The inclusion of
non-equilibrium effects when gradients are large, as performed, for example,
in~\cite{Romatschke:2017vte}, may reduce the discrepancy with the
Planck$+$BK15 data, and we will address this possibility in future work.

A few remarks are worth making before concluding. First, several
  works, for example \cite{bamba2016inflation, Brevik:2017msy,Normann:2016jns,
  maartens1997nonlinear}, have discussed the
  employment of dissipative hydrodynamics with a bulk viscosity to model
  the inflationary phase of the universe's evolution (the latter is the
  only phase of the cosmological evolution we are interested in). In this
  respect, our approach is not novel and has been employed by several
  authors and validated in the literature. Obviously, such a description
  is valid if the local mean-free-path of the particle collisions is much
  smaller than the scale of the system (this is known as the
  Knudsen number criterion~\cite{Rezzolla_book:2013}), which is obviously
  one of our background assumptions. In addition, hydrodynamics has been
  shown to offer a correct quantitative description of systems that are
  not close to local equilibrium, for instance in heavy-ion collision
  physics, e.g., \cite{ATLAS:2015hzw, CMS:2015fgy}, or even when the
  mean-free-path is comparable with the scale of the system, e.g.,
  \cite{Bozek:2010pb, Werner:2010ss}. All of these examples indicate that,
  as an effective theory, hydrodynamics has a validity that is far larger
  than what one would expect by simply considering the Knudsen number
  criterion. Second, while our approach would not be able to describe a
  post-inflationary phase such as reheating, this is not what we are
  dealing with in our manuscript, which concentrates only on the
  evolution of the universe where a hydrodynamical description is
  possible and, indeed, has been employed by numerous authors before
  us. Finally, while it would be extremely interesting to match the end
  of the inflationary phase with one where the particle density is so small
  so as to require the use of a kinetic theory approach, this is well
  beyond the scope of our paper and may be addressed in future work.

\section{Conclusions and~Outlook} 
\label{sec-5}

There is a widespread consensus that the early universe underwent a
phase of quasi de-Sitter expansion, and this is normally modelled by means
of a suitable scalar field and of an associated potential. We have
presented an alternative modelling of the inflationary expansion that is
not based on a scalar field, but that involves, instead, a bulk viscous
cosmological fluid to sustain a quasi de-Sitter expansion in the early
universe. Our model is set in the framework of the generalised causal
theory of hydrodynamics, and by taking into account out-of-equilibrium
effects, it reveals that, if the cosmological fluid possessed a nonzero
bulk viscosity, then a quasi de-Sitter inflation would arise naturally in this
scenario without invoking additional fields and without assuming an
EOS of the type $p=w_0\, e=-e$ relating the equilibrium pressure and the
energy density. Hence, a quasi de-Sitter inflation can be realised purely
with the help of the bulk viscosity and by taking into consideration the
non-equilibrium effects that inevitably arise due to the violation of the
SEC during the accelerated expansion of the early universe.

The model of inflation presented here has several interesting
features. First, while the bulk viscosity provides the necessary negative
pressure required for the accelerated expansion, as a matter of course,
the effective EOS---expressed in terms of the ratio between the effective
pressure and the energy density, $w_{\mathrm{eff}}$, and that embodies the
bulk viscous and non-equilibrium effects---becomes a time-dependent
function describing the time evolution of the inflationary phase. Second, the
evolution of $w_{\mathrm{eff}}$ follows a rather simple and unique
behaviour that reflects a smooth transition from the exact de-Sitter
phase and to a subsequent levelling off to a constant value at later
times. While this behaviour is obtained by numerically solving the
generalized momentum-conservation equation, the functional behaviour of
$w_{\mathrm{eff}}$ can be assimilated as a simple Logistic function, and
hence, the associated timescale for the transition from exact de-Sitter
to the new quasi de-Sitter phase can be estimated accurately. This timescale
is a simple function of the three parameters of the systems
($\bar{\zeta}$, $\alpha$, and $w_0$), and unsurprisingly, it decreases as
the magnitude of the bulk viscosity coefficient is increased up to the
exact de-Sitter value. Third, in contrast with the standard inflationary
scenario where $-1< w_0<1/2$, our bulk viscous model allows for a larger
range of values, namely $-1< w_0<1$, and is well-behaved for $w_0=0$.
Finally, the equivalence between the non-perfect fluid description of
inflation presented here with the scalar field theory of inflation is
maintained also when considering the observational constraints, in
particular when expressing the constraints coming from
Planck2018~\cite{akrami2020planck}, Keck-Array BK15, ACTPol+WMAP, and
SPT3G+WMAP~\cite{Forconi:2021que}, in terms of the standard inflationary
variables like the spectral index of scalar density perturbations $n_s$
and the tensor-to-scalar ratio $r$. We find that these constraints can be
easily satisfied by suitable choices of $\bar{\zeta}$, $\alpha$, and
$w_0$ for the ACTPol$+$WMAP and SPT3G$+$WMAP datasets, while tensions
appear in the case of the Planck$+$BK15 datasets.

The work presented here can be extended and improved in a number of
directions, for example by investigating the possible origin of bulk
viscosity in the early universe and relating it to the mechanism of
particle production~\cite{PhysRevD.53.5483}. Similarly, it would be
interesting to investigate the precise mechanism giving rise to the exit
from the quasi de-Sitter inflationary phase, and this could lead to conducive
conditions for reheating. Similarly, the study of warm inflation recently
suggested in Ref.~\cite{Montefalcone:2022jfw} could also be re-analysed
when framed in the presence of the bulk viscosity and non-equilibrium
effects and, finally, before a kinetic theory approach needs to be
  employed when the the hydrodynamic description used ceases to be
  valid. Some steps in this direction have already been taken, and the
  reheating process of the universe in the presence of the bulk viscosity has
  been addressed by several authors, for example within the warm inflation
  scenario \cite{Zimdahl:1996tg, Zimdahl:1999tn,
    Mimoso:2005bv}. Following their approach, it would be interesting to
  study the reheating process in the contest of the generalised MIS
  theory. We leave all of these investigations to future works.

\vspace{6pt}
\authorcontributions{{Conceptualization S.L. and L.R.; validation, S.L. and L.R.; formal analysis, S.L. and
    L.R.; investigation, S.L. and L.R.; writing---original draft preparation, S.L.;
    writing---review and editing, L.R. and S.L.; visualization, S.L. and
    L.R.; funding acquisition, S.L. and L.R. All authors have read
    and agreed to the published version of the manuscript. } }

\funding{{{S.L. is supported by the Deutsche Forschungsgemeinschaft (DFG)}
 with
grant/40401154. Partial funding also came from the Deutsche
Forschungsgemeinschaft (DFG, German Research Foundation) through the
CRC-TR 211 ``Strong-interaction matter under extreme conditions''---
project number 315477589--TRR 211. } } 

\dataavailability{{Data are contained within the article. } } 

\acknowledgments{\textls[-35]L.R. acknowledges the Walter Greiner
Gesellschaft zur F\"orderung der physikalischen Grundlagenforschung
e.V. through the Carl W. Fueck Laureatus~Chair.}


\conflictsofinterest{{The authors declare no conflicts of interest.} } 

\begin{adjustwidth}{-\extralength}{0cm}
\reftitle{References}




%
\PublishersNote{}
\end{adjustwidth}
\end{document}